\documentstyle[epsfig]{mn2e}
\begin{document}

\title
[Dwarf galaxies in the NGC 1023 Group]
{Dwarf galaxies in the NGC 1023 Group}

\author[Trentham and Tully]
{
Neil Trentham$^{1}$ and R. Brent Tully$^{2}$ \\
$^1$ Institute of Astronomy, Madingley Road, Cambridge, CB3 0HA. \\
$^2$ Institute for Astronomy, University of Hawaii,
2680 Woodlawn Drive, Honolulu HI 96822, U.~S.~A. \\
}
\maketitle

\begin{abstract} 
{ 
We present a compilation of galaxies in the NGC 1023 Group, an accumulation of late-type galaxies at a distance of 10 Mpc.
Members at high and intermediate luminosities were identified from their spectroscopic velocities.
Members at low luminosities were identified from their morphologies on wide-field CCD images.
The faint-end slope is in the range $-1.27 < \alpha < -1.12 $.
There is evidence for two dwarf galaxy populations: one in the halo of NGC 1023 that is dominated by dwarf elliptical galaxies,
and one in the infall region surrounding NGC 1023 that contains mainly dwarf irregular galaxies.  
Similar distinctive populations are observed in the Local Group.

Through our imaging surveys, a picture has emerged of the distributions and types of dwarf galaxies over a wide range of environments.
Of greatest interest is a correlation between dwarf galaxy
population and parent halo mass.
The number of dwarfs per unit mass is constant over the range of halo masses $10^{12} - 10^{14}~M_{\odot}$. 
There are small but significant variations in luminosity functions with location.
These variations can be attributed to differences in the bright and intermediate luminosity populations.
In less evolved regions marked by substantial populations of spirals, galaxies with intermediate luminosity are relatively common, 
and such regions have low mass-to-light ratios.
In more evolved regions where spirals are absent there is relative depletion at intermediate luminosities 
and, overall in such places, mass to light ratios are high.
Statistics are still poor for regions outside of massive halos but the faint end slope of the luminosity function 
is comparable in the field with what is found in the massive halos.
Everywhere that has been studied, the faint end slope is unequivocally shallower than the slope at low masses anticipated by hierarchical clustering theory.
Though the slope is shallow, there is no hint of a cutoff down to an observational limit of $M_R = -10$.
}
\end{abstract} 


\section{Introduction} 

In this paper we continue our study of the dwarf galaxy population in various environments
within the Local Supercluster.  This project was motivated by the observation that in dense
environments like the Virgo (Sandage et al. 1985) and Fornax (Ferguson \& Sandage 1988) Clusters there are many dwarf elliptical galaxies, many of which
are nucleated, yet dwarf galaxies are rare in diffuse spiral-rich environments like the Ursa Major
Cluster (Trentham et al. 2001).  A program was begun to look down to
very faint magnitudes in groups of galaxies of differing size and composition.
If large numbers of dwarf galaxies exist in any environment,
their detection could relieve
the missing satellite problem (Moore et al. 1999, Klypin et al. 1999)
in that environment -- a difficulty for the generally successful ${\Lambda} $CDM galaxy
formation models.

A first look (Trentham \& Tully 2002) suggested that differences do exist between environments
and encouraged a more detailed study.  It soon became clear
that the main limitation in determining the dwarf galaxy properties in nearby groups of galaxies
was not so much in {\it detecting} low-luminosity low-surface brightness (LSB) galaxies as it was
in establishing their distances.  If we could not determine a distance with any accuracy, it would not
be possible to establish group membership.  The emphasis in the project then turned towards
finding methods to distinguish low-surface brightness dwarfs from background galaxies.
Background (luminous) low-surface brightness galaxies are rare, but they exist in enough number
to be a problem.

We found that the most productive way to discriminate group members from LSB background galaxies
is using morphological features.
For example, weak spiral structure or a sharp discontinuity in the light profile are symptomatic
of a distant, luminous galaxy, not a nearby dwarf.  
On the other hand, a galaxy with a central nucleation in an otherwise diffuse and symmetric ellipsoid 
is almost certainly a dwarf of class dE,N.
In order to make these judgments, deep optical images were required.
With large-format mosaic CCDs we
could survey large areas of the sky and obtain meaningful counting statistics
for relatively small groups of galaxies.
The best instrument--telescope combination for the study
was the square-degree MegaCam detector (Boulade et al. 1998) on the 3.6 m Canada-France-Hawaii Telescope (CFHT).

The first two environments we studied were the dense elliptical-rich concentrations of galaxies
around the gE galaxies NGC 5846 and NGC 1407 (Mahdavi et al. 2005, Trentham et al. 2006).  In both cases, we detected many
dwarf elliptical galaxies, similar to those observed in the Virgo Cluster.  The luminosity function
had a logarithmic slope $\alpha \sim -1.35$ over the magnitude range $-17<M_R<-12$, subtly but significantly
steeper than the values $-1.1 < \alpha < -1.3$ seen in the Local Group (van den Bergh 2000) and in the field 
(Blanton et al. 2003, Norberg et al. 2002).

The next group given attention was around the luminous Sb galaxy
NGC 5371 (Tully \& Trentham 2008).  This group has a considerably larger spiral fraction than either of the
groups discussed in the previous paragraph so might have provided a sample in an environment
common to most galaxies.
It turned out that there is a significant dwarf galaxy population in this group but centered
on the interacting S0 pair NGC 5353/4.
We thought we would be studying a spiral-rich, presumably unevolved, group, but found a
more complicated situation.   The group has a red, dead core that had recently received
an influx of gas-rich systems.

\begin{figure}
\begin{center}
\vskip-4mm
\psfig{file=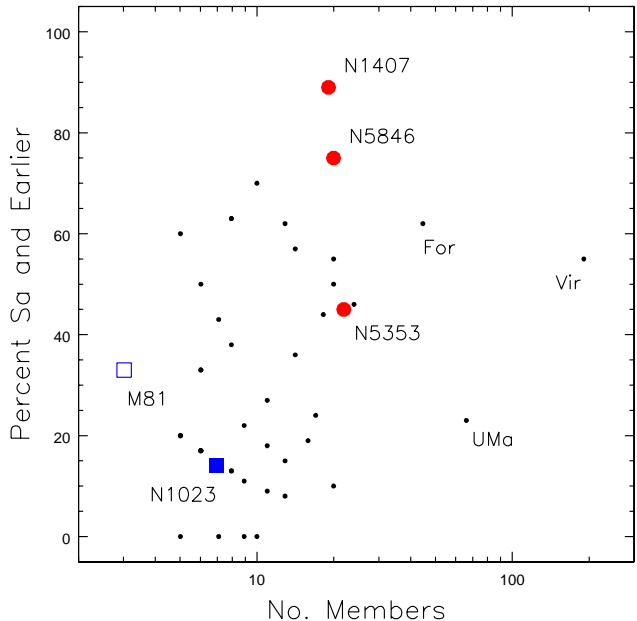, width=8.65cm}
\end{center}
\vskip-3mm
\caption{     
Demographics of nearby groups.  The large colored symbols indicate the compositions of groups in this series of studies based on CFHT MegaCam observations.
The number of members with $M_R<-18$ is plotted against the percentage of these galaxies typed Sa and earlier. 
The small black dots represent 
groups with at least 5 members brighter than $M_B = -16$ and within 25~Mpc from Tully (2005).
}
\label{allgroups}
\end{figure}

A characterization of nearby clusters and groups of galaxies is shown in
Figure~\ref{allgroups}.  The abscissa gives a measure of the group richness and the
ordinate gives a measure of the degree of evolution, marked by the transformation of
galaxies from gas-rich to gas-poor (see the discussion in Kormendy
\& Bender 1996).  The small black symbols are drawn from a complete
high latitude sample of groups within 25 Mpc discussed by Tully (1987, 2005).  The first two
groups that we studied, those around NGC 1407 and NGC 5846, are both modest in richness
but among the most highly evolved of local structures.  The group around NGC 5353 was selected
for its spiral content but turns out to be a composite of evolved and unevolved components.
In a companion program, Chiboucas, Karachentsev, and Tully (2009) have completed a study
of the sparse, spiral-dominant M81 Group.

This paper extends the project with a study of the group around NGC 1023.
This is a spiral-rich group of low enough density that it is representative
of the environments in which most galaxies in the Universe reside.  At 10 Mpc (Tully et al. 2009),
it is considerably closer than the other groups, except M81.  Therefore, we might expect to probe
further down the luminosity function.  Being so nearby, it was necessary to image
a substantial area of the sky.   As the group is so sparse, it can be anticipated that the confusion
between group members and contaminants might be intimidating.  The groups studied earlier
gave us training to tackle this more difficult case.

A concern of studies of poor groups such as this one is that of counting statistics.
The group contains only a few luminous galaxies, and this leads to severe Poisson
uncertainties.  This is the inevitable consequence of an attempt to study an environment
common to most galaxies.  Spiral-rich environments are typically
diffuse groups.  Large accumulations
of spiral galaxies like the Ursa Major Cluster (Tully et al. 1996, Trentham et al. 2001) are rare.  The Ursa Major Cluster
is sufficiently distant that we cannot probe the luminosity function down to the the faint levels studied
in this paper.

Another interesting comparison will be with the Local Group (van den Bergh 2000).  This environment is the one
where the luminosity function has been extended to the faintest known representatives of galaxies in
the Universe (Koposov et al. 2008).
The Local Group consists of
two large halos in the process of virializing, embedded in a common infall region.  The
luminosity function of the ensemble is numerically dominated by the populations in the
virializing regions.  The statistics offered by the Local Group are insufficient to reveal a 
difference in the luminosity functions of the infall and collapsed regions.  The observations
discussed here include the collapsed core and infall region around NGC 1023, so provide some
enlightenment regarding the luminosity function at the extremely low densities of the infall domain.

In Section 2 we describe the NGC 1023 environment and in Section 3 we describe our observations.
In Section 4 we present the sample.  In Section 5
we discuss the results of a dynamical analysis and in Section 6 we present the luminosity function.
In Section 7 we place these results into a consistent framework and Section 8 is a summary.

\section{The Choice of Survey Area}

\begin{figure}
\begin{center}
\vskip-4mm
\psfig{file=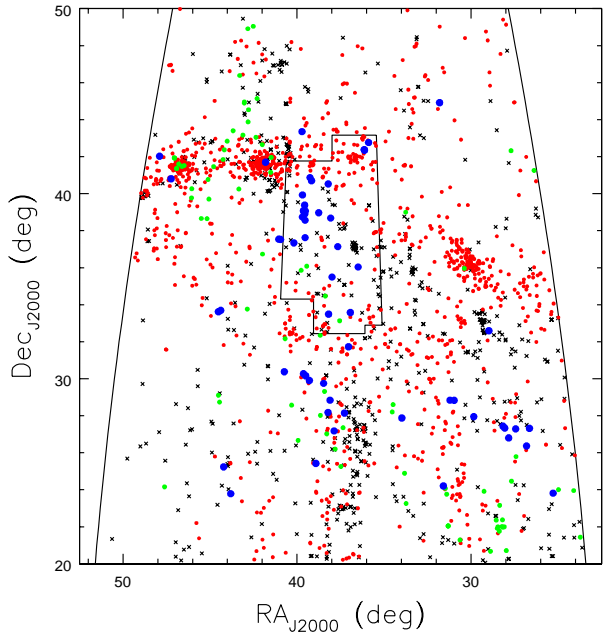, width=8.65cm}
\end{center}
\vskip-3mm
\caption{
Distribution of galaxies in a wide 
area around the survey region.  The positions of galaxies are plotted if they have known velocities less than 12,000 ${{\rm km}\,{\rm s}^{-1}}$,
right ascensions between 1h30 and 
3h30, and declinations between +20 and +50.  Galaxies with heliocentric velocities less than 1100 
${{\rm km}\,{\rm s}^{-1}}$\ are indicated by 
blue symbols, cases with velocities between 1100 and 3500 ${{\rm km}\,{\rm s}^{-1}}$\ are 
in green, those with velocities between 3500 and 7000 ${{\rm km}\,{\rm s}^{-1}}$\ are in red, and more distant
objects to the 12,000 ${{\rm km}\,{\rm s}^{-1}}$\ limit are indicated by black crosses.  The survey region is 
within the irregular box at the center.  There are 4 clusters within
the background filament that passes through the top of the survey region.  Abell 426 (Perseus Cluster) is at the left edge of the figure, 
AWM7 is just to the
upper left of the survey region.  Abell 347 is in the upper right corner of the survey region.  Abell 262 is to the right of the survey region.  These 
clusters all lie in the range 5,000--6,000 ${{\rm km}\,{\rm s}^{-1}}$.
}
\label{wideview}
\end{figure}

\begin{figure}
\begin{center}
\vskip-4mm
\psfig{file=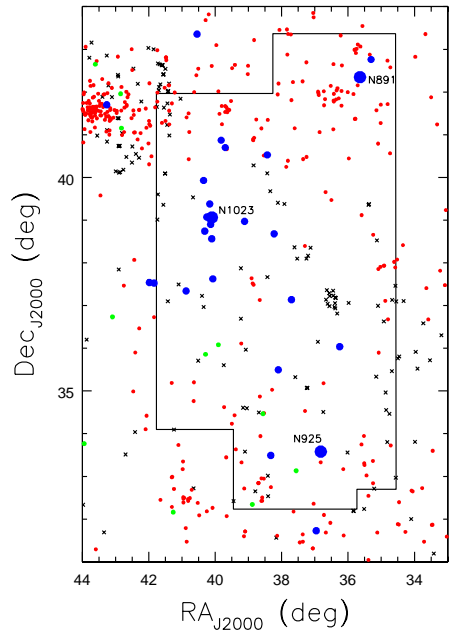, width=8.65cm}
\end{center}
\vskip-3mm
\caption{
Zoom to the center of Fig.~\ref{wideview}.  The galaxies identified by the large 
blue symbols have heliocentric velocities between 400 and 1000 ${{\rm km}\,{\rm s}^{-1}}$\ and are
suspected to be members of a bound group.  The 3 most luminous galaxies are identified.  The background cluster AWM7 to the upper left of the survey region
lies at a heliocentric velocity of 5168~${{\rm km}\,{\rm s}^{-1}}$.  The cluster in the vicinity of NGC~891 is Abell 347 
at 5940~${{\rm km}\,{\rm s}^{-1}}$.
}
\label{surveyregion}
\end{figure}

We wanted to study a region with a concentration of spirals but otherwise as isolated as possible.
The target should be nearby so the luminosity function could be accessed to faint levels.
As a consequence of the topology of local structure, most nearby groups are confused.  Unless accurate
distances are available, problems arise from contamination between separate structures in the line of sight.

The group around NGC 1023 is reasonably well disposed with regard to these concerns.
Figure~\ref{wideview} provides the big picture.  Symbol color, shape, and size provides depth discrimination
for galaxies with known velocities in the region of interest.  The dominant structure, coded red,
is the Perseus--Pisces filament (Haynes \& Giovanelli 1988) at $\sim 5000$~${{\rm km}\,{\rm s}^{-1}}$.  Our interest is
in a grouping of the objects in blue at $\sim 600$~${{\rm km}\,{\rm s}^{-1}}$.
Figure~\ref{surveyregion} is a blow up of the central part of Fig.~\ref{wideview}.  The blue symbols
cluster around NGC~1023, the brightest galaxy in the region.  The CFHT MegaCam survey was undertaken
in the area bounded by the irregular box.  The second and third brightest galaxies in the region,
NGC 891 and NGC 925, are included within the survey.

We wanted to study a volume rich in spirals but it must be noted that NGC 1023 is a lenticular galaxy.
However, all the other 26 galaxies with velocities less than 1000~${{\rm km}\,{\rm s}^{-1}}$\ plotted in Fig.~\ref{surveyregion}
are spirals or magellanic irregulars.

\begin{figure}
\begin{center}
\vskip-4mm
\psfig{file=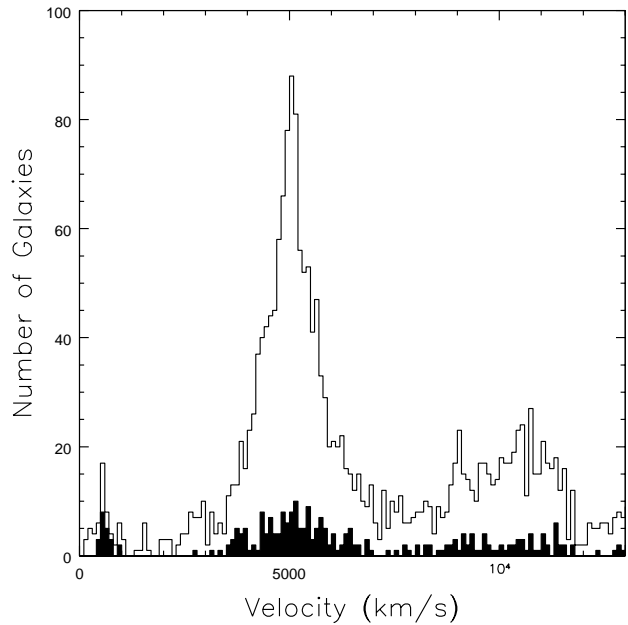, width=8.65cm}
\end{center}
\vskip-3mm
\caption{
Histogram of velocities of galaxies in the area about the survey.  Open histogram: all galaxies in the region illustrated in Fig.~\ref{wideview}.  Filled 
histogram: only galaxies within the CFHT MegaCam survey region. The velocity regimes of the blue, red, and black symbols in the two previous figures are 
chosen
to capture the structure in the 3 histogram peaks, at less than 1000~${{\rm km}\,{\rm s}^{-1}}$, 
$\sim 5000$~${{\rm km}\,{\rm s}^{-1}}$, and $\sim 10,000$~${{\rm km}\,{\rm s}^{-1}}$, respectively.  The green symbols locate
objects between the local and 5000~${{\rm km}\,{\rm s}^{-1}}$\ structures.
}
\label{vhist}
\end{figure}

Histograms of heliocentric velocities are given in Figure~\ref{vhist}; open for the larger region of
Fig.~\ref{wideview} and filled for the restricted region of the CFHT survey.  Peaks in the distribution
occur at 600, 5000, and 10,000~${{\rm km}\,{\rm s}^{-1}}$.  In the immediate field of the survey, there are no known galaxies
between 1000 and almost 3000~${{\rm km}\,{\rm s}^{-1}}$.  If one views the distribution of galaxies in three dimensions,
it is found that a continuous filament runs from the Local Group through the NGC 1023 Group and all the
way to the Perseus--Pisces structure.  Members of this connecting filament are seen in green in
Figs. \ref{wideview} and \ref{surveyregion}. There is possible contamination from this structure in the extreme bottom--left
corner of the survey.  The area of the survey includes most of the known galaxies associated with
NGC 1023 and is otherwise devoid of galaxies until the distance of the Perseus--Pisces complex.

It is to be acknowledged that the Galactic latitude of the NGC 1023 Group is uncomfortably low at
$b \sim -20$.  However, obscuration is low across the survey region at $A_R \sim 0.15-0.20$ (Figure 5). 
The situation is not ideal, but the NGC 1023 Group is
the most suitable target for our study.

\begin{figure}
\begin{center}
\vskip-4mm
\psfig{file=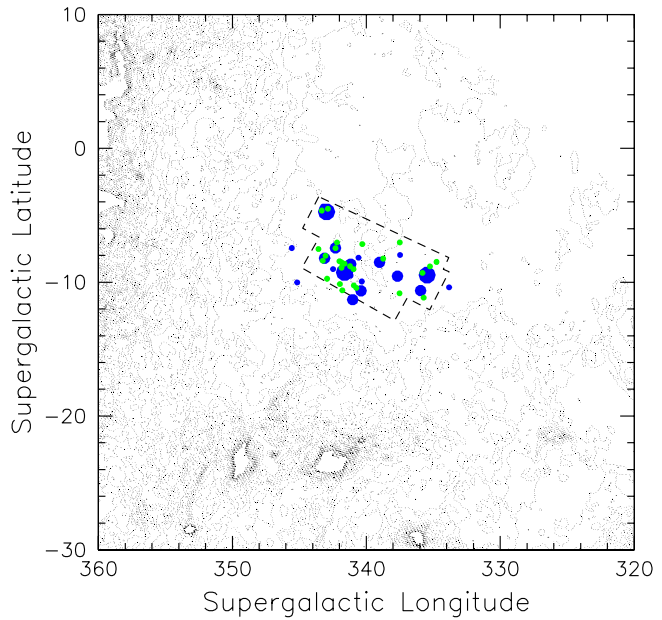, width=8.65cm}
\end{center}
\vskip-3mm
\caption{
Proximity of Galactic obscuration.  The outline of the survey region is 
superposed on a map of Galactic reddening given by Schlegel et al. (1998).  The $R$ band 
absorption is in the range 0.15--0.20 mag, never more than 0.25 mag.  The obscuration 
increases rapidly beyond the survey region toward the Galactic equator, 
coincident with the left edge of the figure, and toward the high latitude 
complex in Orion, toward the bottom of the figure.  The blue circles locate the 
galaxies with known velocities within 400--1000~${{\rm km}\,{\rm s}^{-1}}$\ suspected to be within the 
bound group.  The green symbols identify candidates found in the CFHT MegaCam 
survey}
\label{sglb}
\end{figure}

\section{Observations}

\subsection{The Imaging Survey}

Observations were made with the MegaCam mosaic CCD camera (Boulade et al. 1998) on the 
Canada--France--Hawaii 3.6 m telescope
on Mauna Kea.  Exposures were taken in queue mode under approximately photometric condition in July-December
2003.  The median seeing was 0.7 arcseconds.  The MegaCam detector is a mosaic of 36 CCDs, providing a field
of view of 60$^{\prime}$ x 60$^{\prime}$.  Images were made with a dither of half a field of view of the camera
so that gaps were filled in and so that every piece of the sky was observed twice, allowing us to establish
that the data were taken under photometric conditions.  The data were reduced and calibrated using the
Terapix (Bertin et al. 2002) pipeline.
A total of 109 x 5 minute exposures were taken, all using the Gunn $r$-band.  As in the previous paper
in this series (Tully \& Trentham 2008) magnitudes have been transformed to the Cousins $R$ system
defined by Landolt (1992).
A total area of 60.7 deg$^2$ was covered down to a limiting surface brightness of 1$\sigma$ = 27
mag arcsec$^2$ within an aperture of FWHM = 2 arcsec.

In an earlier study (Trentham \& Tully 2002, hereafter TT02) we presented deep observations
of the core of the NGC 1023 group taken with the mosaic CCD SuprimeCam detector on the 8.2 m
Subaru Telescope.  The current survey of 60 times the area covers much of the infall region
around the central galaxy.  The large angular size of this survey is also useful in establishing
the morphological characteristics and density
of field LSB galaxies.  This is particularly important in the current study because
the NGC 1023 group is very diffuse and counting statistics are a problem.  Therefore
minimizing other sources of uncertainty, such as contamination in the sample from field LSB galaxies,
is particularly important.

\subsection{Subaru spectroscopic observations}

\begin{table}
\caption{Results of Subaru Spectroscopy}
{\vskip 0.55mm} {$$\vbox{ \halign {\hfil #\hfil && \quad \hfil #\hfil \cr
\noalign{\hrule \medskip}
Confirmed group members\cr
\noalign{\medskip \hrule \medskip}
ID~~~~~$V_h$~km/s\cr
11~~~~$740 \pm ~43$\cr
23~~~~$448 \pm ~57$\cr
42~~~~$455 \pm 187$\cr
\noalign{\medskip \hrule \medskip}
Background\cr
\noalign{\medskip \hrule \medskip}
2MASSXJ02392281+4034032~~$23781 \pm ~61$\cr
2MASSXJ02401416+3820499~~~$9405 \pm ~38$\cr
\cr
  \hline
\noalign{\smallskip \hrule}
\noalign{\smallskip}\cr}}$$}
\end{table}

Optical spectra were obtained for a small number of galaxies in the area of NGC 1023 as a secondary
program.  Observations were made with the Faint Object Camera and Spectrograph (FOCAS; Saito et al. 2003) at the Cassegrain
focus of Subaru Telescope over the nights 21--23 November, 2006.  Kristin Chiboucas and Andisheh Mahdavi
participated in the observations as co-investigators in the primary program (discussed elsewhere).
The instrumental set-up involved the 600-450 grism providing a dispersion of $0.37^{\prime \prime}$~pix$^{-1}$
over the spectral range 3765--5260 \AA.  A $2^{\prime \prime}$ slit was used to maximize the signal 
received from low surface brightness candidates.

Five targets were successfully observe.  Two of these were high surface brightness early-type systems
that we suspected are giant galaxies in the background but, on morphological grounds, could be small 
nearby ellipticals.  It was not a surprise to find that these two objects are indeed in the background.
The other three targets were suspected to be group members and were confirmed to be so from their 
observed velocities.  They are identified by the ID numbers introduced in Table~2.  In all cases, 
the only identifiable spectral features are absorption lines.
Candidate ID 42 is extremely faint and the velocity assignment is uncertain.  

\subsection{Neutral Hydrogen observation}

The extremely large, low surface brightness object Candidate ID 38 was observed in the 21cm line of HI
by H\'el\`ene Courtois on 2007 December 2 with the 100m Green Bank Telescope.  The target was not detected
at the level of 1~Jy~km~s$^{-1}$.

\section{Membership considerations and the catalog}

\begin{table*}
\caption{The NGC 1023 Group Sample}
{\vskip 0.55mm} {$$\vbox{ \halign {\hfil #\hfil && \quad \hfil #\hfil \cr
\noalign{\hrule \medskip}
ID  & Name &  Type & Rating & $V_h$ & $\alpha$ (J2000) & $\delta$ (J2000) & $s$ &  ID$_{\rm TT02}$ & $R$ & $M_R$ 
                      &\cr
    &            &      &    &  km/s    &  &             &        &        &         &    &   &          &\cr
\noalign{\smallskip \hrule \smallskip}
\cr
   1  &  NGC 1023  &  S0      &  0  &   637  & 02 40 24.0  &   39 03 48  & 0    &  1 &   7.83  &  -22.33 &\cr
   2  &  NGC 891   &  Sb      &  0  &   528  & 02 22 33.4  &   42 20 57  & 279  &         &  7.86 &  -22.31 &\cr
   3  &  NGC 925   &  Sd      &  0  &   553  & 02 27 16.9  &   33 34 45  & 368     &         &  9.55  &  -20.65 &\cr
   4  &  NGC 1003  &  Sd      &  0  &   626  & 02 39 16.9  &   40 52 20  & 109     &    &  10.83  &  -19.36 &\cr
   5  &  NGC 1058  &  Sc      &  0  &   518  & 02 43 30.0  &   37 20 29  & 110     &         &  10.85  &  -19.32 &\cr
   6  &  NGC 949   &  Sb      &  0  &   609  & 02 30 47.6  &   37 08 38  & 163     &    &  10.92  &  -19.24 &\cr
   7  &  NGC 959   &  Sdm     &  0  &   597  & 02 32 24.0  &   35 29 44  & 235     &    &  11.79  &  -18.40 &\cr
   8  &  IC 239    &  Scd     &  0  &   903  & 02 36 27.9  &   38 58 12  & 46      &         &  12.75  &  -17.44 &\cr
   9  &  UGC 2034  &  Im      &  0  &   578  & 02 33 42.9  &   40 31 41  & 116  &         &  13.45  &  -16.70 &\cr
  10  &  UGC 2023  &  Sc      &  0  &   589  & 02 33 18.2  &   33 29 28  &346&    &  13.59  &  -16.67 &\cr
  11  &  UGC 2165  &  dE,N    &  0  &   740  & 02 41 15.5  &   38 44 36  &22 &  2  &  13.70   &  -16.46 &\cr
  12  &  UGC 2126  &  Sc      &  0  &   713  & 02 38 47.1  &   40 41 55  & 100  &    &  13.98  &  -16.20 &\cr
  13  &  UGC 2157  &  Sdm     &  0  &   488  & 02 40 25.1  &   38 33 48  & 30      &  3  &  13.99  &  -16.15 &\cr
  14  &  UGC 1807  &  Im      &  0  &   629  & 02 21 13.4  &   42 45 46  &306&    &  14.38  &  -15.83 &\cr
 15  &  NGC 1023A &  dI      &  0  &   743  & 02 40 37.7  &   39 03 27  & 3    &  4 &  14.52  &  -15.64 &\cr
  16  &  UGC 2014  &  Sdm     &  0  &   565  & 02 32 54.0  &   38 40 50  & 91&    &  14.50  &  -15.63 &\cr
  17  &  UGC 1865  &  Sdm     &  0  &   580  & 02 25 00.2  &   39 02 16  &   260 &   &  14.78  &  -15.41 &\cr
  18  &            &  dE/I    &  2  &        & 02 37 18.5  &   38 56 00  &37 &       &  14.92  &  -15.28 &\cr
 19  &  NGC 1023C &  dE/I    &  0  &   903  & 02 40 39.6  &  39 22 47   & 19  &5 & 16.31 & -13.84 &\cr
  20  &  NGC 1023B &  dI      &  0  &   593  & 02 41 00.0  &   39 04 19  & 7    &  8 &  16.35  &  -13.82 &\cr
  21  &            &  dE/I    &  3  &        & 02 43 01.6  &   37 59 27 &72&    &  16.37  &  -13.77&\cr
  22  &  NGC 1023D &  dE/I    &  0  &   695  & 02 40 33.0  &   38 54 01  & 10 & 6  &  16.51  &  -13.65 &\cr
  23  &            &  dE,N    &  0  &   448  & 02 40 17.0  &   37 37 34  &86  &  7&  16.66  &  -13.49 &\cr
  24  &            &  dE,N    &  3  &        & 02 37 36.0  &   34 46 06  &260&   &  16.72  &  -13.46 &\cr
  25  &            &  dE      &  3  &        & 02 21 12.1  &   42 21 51  &291&   &  16.85  &  -13.27 &\cr
  26  &            &  dI      &  2  &        & 02 45 50.0  &   39 57 12  &82&   &  17.24  &  -12.99 &\cr
  27  &            &  dE      &  2  &        & 02 31 29.3  &   40 37 12  &138&   &  17.25  &  -12.87 &\cr
  28  &            &  dI      &  3  &        & 02 27 35.8  &   38 53 46  &150&  &  17.29  &  -12.82 &\cr
  29  &            &  dI      &  3  &        & 02 37 27.0  &   39 22 46  &39&   &  17.65  &  -12.52 &\cr
  30  &            &  dI      &  2  &        & 02 22 55.2  &   42 42 42  &292&  &  17.98  &  -12.20 &\cr
  31  &            &  dI      &  2  &        & 02 33 47.5  &   40 31 07  &115&   & 18.18  &  -11.97 &\cr
  32  &            &  dI      &  2  &        & 02 46 48.4  &   38 32 53  &81&    & 18.22  &  -11.95 &\cr
  33  &            &  dE      &  3  &        & 02 37 39.4  &   38 36 02  &42&    & 18.39  &  -11.79 &\cr
  34  &            &  dI      &  3  &        & 02 37 18.4  &   41 36 10 &156&    &  18.43  &  -11.75 &\cr
  35  &            &  dE      &  1  &        & 02 37 31.1  &   39 37 48  &48&    &  18.45  &  -11.73 &\cr
  36  &            &  dE      &  2  &        & 02 40 29.9  &   40 53 37 &110&    &  18.78  &  -11.41 &\cr
  37  &            &  dE,N    &  2  &        & 02 45 09.4  &   38 56 37  &56&    &  18.88  &  -11.29 &\cr
  38  &            &  VLSB    &  1  &        & 02 39 2l.0  &   39 26 17   &26&  9  &  18.90  &  -11.25 &\cr
  39  &            &  dE      &  3  &        & 02 35 04.3  &   33 03 12 &367&   &  19.22  &  -11.02 &\cr
  40  &            &  dE      &  3  &        & 02 27 20.4  &   33 57 22 &347&   &  19.19  &  -11.00 &\cr
  41  &            &  dE      &  1  &        & 02 38 17.1  &   40 53 53 &113&    &  19.24  &  -10.93 &\cr
  42  &            &  dE      &  0  &   455  & 02 41 23.9  &   39 55 46  &53& 10 &  19.25  &  -10.93 &\cr
  43  &            &  dE,N    &  2  &        & 02 43 24.5  &   37 44 26  &87&    &  19.30  &  -10.84 &\cr
  44  &            &  dE      &  2  &        & 02 37 30.9  &   39 47 45  &55&    &19.38& -10.81 &\cr
  45  &            &  dE/I    &  3  &        & 02 20 50.9  &   36 27 39 &283&    &19.36& -10.80 &\cr
  46  &            &  dE/I    &  3  &        & 02 24 02.4  &   33 40 47 &382&   &19.45& -10.76 &\cr
  47  &            &  dE/I    &  3  &        & 02 28 59.3  &   37 01 45 &183&   &19.44& -10.70 &\cr
  48  &            &  dI      &  2  &        & 02 21 35.5  &   33 22 37 &415&   &19.51& -10.70 &\cr
  49  &            &  dI      &  2  &        & 02 42 39.6  &   41 22 45 &141&   &19.58& -10.66 &\cr
  50  &            &  dE/I    &  3  &        & 02 40 30.1  &   38 29 39  &34  & 12 &  19.52  &  -10.62 &\cr
  51  &            &  dE      &  3  &        & 02 33 23.5  &   34 17 05 &300&   &19.58& -10.60 &\cr
  52  &            &  dE/I    &  3  &        & 02 34 54.0  &   33 04 47&366 &   &19.77& -10.47 &\cr
  53  &            &  dE      &  2  &        & 02 38 33.4  &   37 05 53&120 &   &19.72& -10.42 &\cr
  54  &            &  dE,N    &  3  &        & 02 43 56.5  &   37 24 26&108 & &19.79     &  -10.37 &\cr
  55  &            &  dE/I    &  3  &        & 02 29 33.3  &   41 57 08&211 &  &20.11&  -10.07 &\cr
\cr
\noalign{\smallskip \hrule}
\noalign{\smallskip}\cr}}$$}
\end{table*}

\begin{table*}
{\vskip 0.55mm} {$$\vbox{ \halign {\hfil #\hfil && \quad \hfil #\hfil \cr
\noalign{\hrule \medskip}
ID  & Name &  Type & Rating & $V_h$ & $\alpha$ (J2000) & $\delta$ (J2000) & $s$ &  ID$_{\rm TT02}$ & $R$ & $M_R$
                      &\cr
    &            &      &    &  km/s    &  &             &        &        &         &    &   &          &\cr
\noalign{\smallskip \hrule \smallskip}
\cr
  56  &            &  dE/I    &  3  &        & 02 20 02.9  &   34 26 39&374 &  &20.21&  -10.03 &\cr
  57  &            &  dE/I    &  3  &        & 02 29 05.8  &   37 39 05&159 &  &20.14&  -10.01 &\cr
  58  &            &  dE/I    &  3  &        & 02 34 15.6  &   32 52 22&379&&  20.33      &  -9.95 &\cr
  59  &            &  dE      &  3  &        & 02 38 39.4  &   40 15 04 &74& & 20.44      &  -9.72 &\cr
  60  &            &  dI      &  2  &        & 02 34 15.8  &   32 52 22&379& & 20.75      &  -9.53 &\cr
  61  &            &  dE/I    &  3  &        & 02 39 59.5  &   38 24 07 &40&23&21.05  &  -9.07 &\cr
  62  &            &  dE/I    &  2  &        & 02 36 23.3  &   40 40 12&107& & 21.11      &  -9.36 &\cr
  63  &            &  dE/I    &  3  &        & 02 31 29.5  &   40 27 34&132& & 21.10  &  -9.04 &\cr
  64  &            &  dE/I    &  3  &        & 02 32 39.9  &   41 42 21&181& & 21.19      &  -8.99&\cr
  65  &            &  dE/I    &  3  &        & 02 41 16.6  &   39 23 49 &22&19&21.37    &  -8.79 &\cr
\cr
\noalign{\smallskip \hrule \smallskip}
  *   & UGC 2259   & Sdm      &  0  &   583  & 02 47 55.4  &   37 32 18 &127& &12.8~  & -17.4~ &\cr 
  *   & UGC 2172   & Im       &  0  &   544  & 02 42 10.8  &   43 21 18 &258& &13.6~  & -16.7~ &\cr
  *   & UGC 1924   & Scd      &  0  &   598  & 02 27 49.9  &   31 43 35 &464& &14.2~  & -16.0~ &\cr
  *   & UGC 2254   & Im       &  0  &   578  & 02 47 21.7  &   37 31 30 &123& &16.~~  & -14.~~ &\cr
  *   & PGC3096870 & dI       &  0  &   523  & 02 53 03.9  &   41 42 00 &216& &17.~~  & -13.~~ &\cr 
\noalign{\smallskip \hrule}
\noalign{\smallskip}\cr}}$$}
\end{table*}

Within the area of this survey, there are about 100,000 galaxies detected by the Terapix pipeline.
Roughly 99.9\% are
in the background.  We are interested in the 0.1\% of low luminosity galaxies that
are in the group.  These normally have the distinguishing property that they have
low surface brightnesses.

We performed an initial cut from the 100,000 galaxies that are detected to a much smaller number for more detailed study.
Following previous studies, galaxies
are considered with inner and outer concentration parameters (ICP and OCP)
\begin{equation}
ICP = R(4.4~{\rm arcsec}) - R(2.2~{\rm arcsec}) < -0.7
\end{equation}
\begin{equation}
OCP = R(12~{\rm arcsec}) - R(6~{\rm arcsec}) < -0.4.
\end{equation}
Faint galaxies ($R<20$) are given consideration if they extend across more than 2 arcsec$^2$ with surface brightness $\mu_R < 24.5$~mag~arcsec$^{-2}$.
These criteria reduce the number of objects in the sample from 100,000 to about 250.

The sample derived from these criteria contains most group dwarfs (the exceptions being compact ellipticals like M32; de Vaucouleurs 1961) but it admits background galaxies.
Some are distinguished 
on the basis of morphology.  Background galaxies that pass the surface brightness test 
frequently have central high surface brightness
components, or hints of spiral structure, or indications of tidal interactions.  We qualitatively rate all the candidates as (1) probable members, (2) 
possible members, (3) plausible members, and (4) probable background.  Our terminology has turned out to be 
conservative in our previous studies of this series.  Two-point correlation analyses and
spectroscopic follow-ups provided evidence that galaxies 
rated 1 and 2 (probable/possible) were almost always group members, galaxies rated 3 (plausible)
were members about 50\% of the time, and galaxies rated 4 were background.
While these numbers provide a useful guide for the current project, the fact that we may be dealing with later type dwarf galaxies means
that we need an internally consistent way of assessing the ratings.  The strategy described below in which we use galaxies whose positions near
NGC 1023 suggest a high probability of membership to construct a training set reflects this need.

There were additional pieces of information we needed to take into account when compiling the sample.
Point sources as faint as $R \sim 24$ resolve in each 
5 minute exposure so, at distance modulus 30.0, individual very bright supergiant stars are seen and
current star formation regions can be identified.  The problem is with systems with no young stars.
In the earlier studies of the NGC 5846, NGC 1407, and NGC 5353/4 groups the surface densities of the dwarf
populations were high and the groups had minimal line-of-sight contamination.
In the current case of the NGC 1023 Group, the line-of-sight confusion only becomes severe with the
profusion of objects in the Perseus--Pisces filament at 60 Mpc.  Indeed, after we had established a
tentative list of candidates based on the criteria used previously it was found that, while the galaxies
with membership probability ratings 1 or 2 (probable/possible) were strongly correlated with the position
of NGC 1023, the galaxies rated 3 (plausible members) were very poorly correlated with NGC 1023 and
were better correlated with the Perseus--Pisces structure at 5,000 km~s$^{-1}$.
A galaxy of given intrinsic properties in the Perseus--Pisces filament
is offset with respect to its appearance in the NGC 1023 Group by 4 magnitudes in brightness and a
factor 6 in size.  A dE galaxy in the NGC 1023 Group with $M_R \sim -13$ to $-15$ has no self-similar counterpart 4 mag brighter
at $M_R \sim -17$ to $-19$ in the Perseus--Pisces filament. 

The problem is to distinguish a dE at $M_R \sim -9$ in the NGC 1023 Group 
from a dE at $M_R \sim -13$ at 5,000 km~s$^{-1}$.
We can be confident that galaxies with appropriate velocities or rated 1--2 
are in the nearby group but the location of so many marginal candidates in the Perseus--Pisces filament suggests that we have many
near--background contaminants.  Using the high confidence members as a training set, we
used stricter morphological screening to arrive at the list of candidates given in Table~2.  We
consider that the list is comprehensive at the level of 90\% down to $M_R \sim -11$ (the brightest 40 
candidates).  Faintward, the list is incomplete and probably confused with contaminants.  Candidates 
can easily be identified from our images at $M_R \sim -8$ and even fainter but they cannot be confirmed 
as group members.

\begin{figure}
\begin{center}
\vskip-4mm
\psfig{file=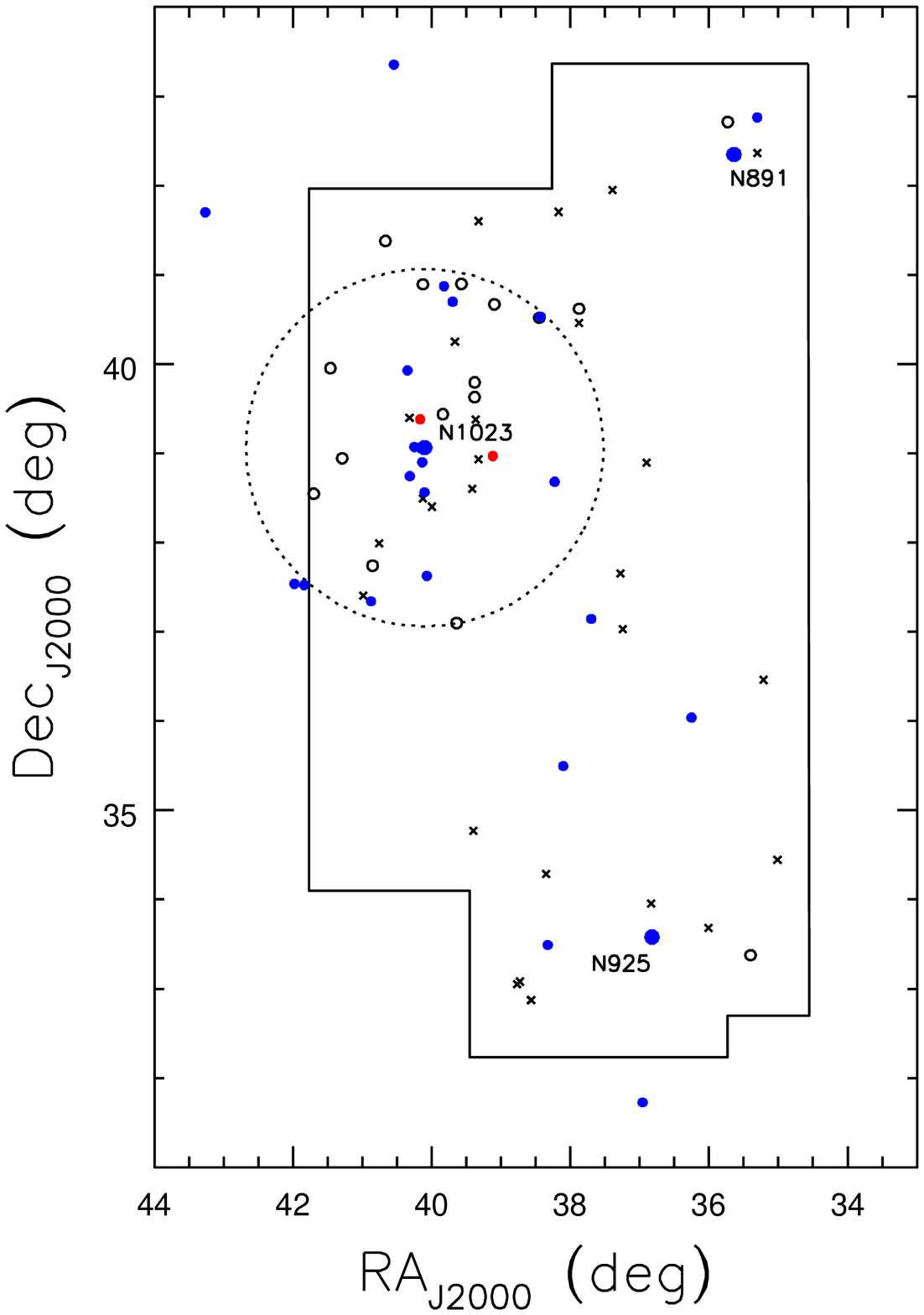, width=9.5cm}
\end{center}
\vskip-3mm
\caption{
Group members and candidate group members of the extended NGC~1023 Group.  Galaxies with observed heliocentric velocities in the interval 400--800 km
s$^{-1}$ are
indentified by blue circles.  Two galaxies with velocities $\sim 900$ km s$^{-1}$ are represented in red.  The domain of the CFHT MegaCam survey is 
contained within
the irregular rectangular outline.  Candidates found in the survey that are considered almost certain group members are identified by open circles.  
Candidates that are probable but less certain members are located by crosses.  The three dominant galaxies 
are identified.  The circle of radius 2 degrees (350 kpc) 
approximates the outer surface of the second turnaround sphere around NGC~1023.
}
\label{cand-all}
\end{figure}

Table~2 provides information on 65 candidates within the wide-field imaging survey region.  These candidates either have known velocities consistent with 
the group (hence given a membership probability rating 0) or they have membership ratings 1-3.
In the table we list for each galaxy its type, rating, coordinates, distance $s$ from NGC 1023 in arcminutes,
ID in TT02, apparent $R$ magnitude, and absolute magnitude.
The absolute magnitude $M_R$ was calculated from the apparent magnitude $R$ using the equation
\begin{equation}
M_R = R - A_R - {\rm DM}
\end{equation}
where $A_R$ is the Galactic extinction (Schlegel et al. 1998) and DM is the distance modulus, 
assumed to be 30.0 for the NGC 1023 Group.

The magnitudes presented in the table for all but the brightest galaxies are aperture magnitudes 
where the aperture size is set
to equal a radius larger than that at which the galaxy blends into the sky.
This method was motivated by the comparison between MegaCam and deep SuprimeCam data
described by Trentham et al. (2006).  For the brightest 7 galaxies which are saturated in the
MegaCam images, magnitudes are taken from the compilation in the Homogenized Photometry catalog
found in the Extragalactic Distance Database\footnote{http://edd.ifa.hawaii.edu} (Tully et al.
2009).  An $R$ band magnitude is directly available for NGC 925 but in the other cases the $R$
magnitude is inferred from the $I$ magnitude assuming $R-I = 0.40$.  These luminous galaxies are
affected by internal dust obscuration.  Corrections to face-on orientation are made following 
the recipe in Tully et al. (1998).  The $R$ magnitudes given in Table~2 include corrections
for internal obscuration for the brightest seven galaxies; such corrections are negligible for
the fainter galaxies.   

The five unnumbered galaxies at the bottom of Table~2 identify systems with known velocities that give
an association with the NGC 1023 Group but which lie outside the boundaries of the CFHT MegaCam survey.

\section{Group Structure and Dynamics}

\begin{figure}
\begin{center}
\vskip-4mm
\psfig{file=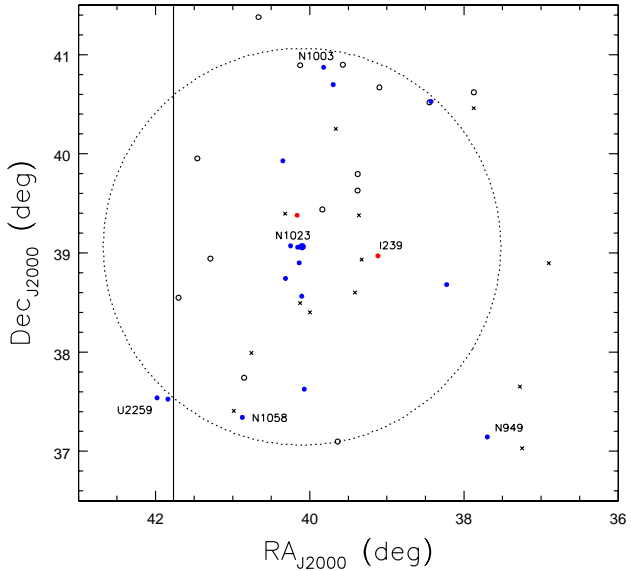, width=8.65cm}
\end{center}
\vskip-3mm
\caption{
Zoom of Fig.~6 into the second infall (virial) region centered on NGC~1023.  The most prominent galaxies are identified.
}
\label{cand-r2t}
\end{figure}

\begin{figure}
\begin{center}
\vskip-4mm
\psfig{file=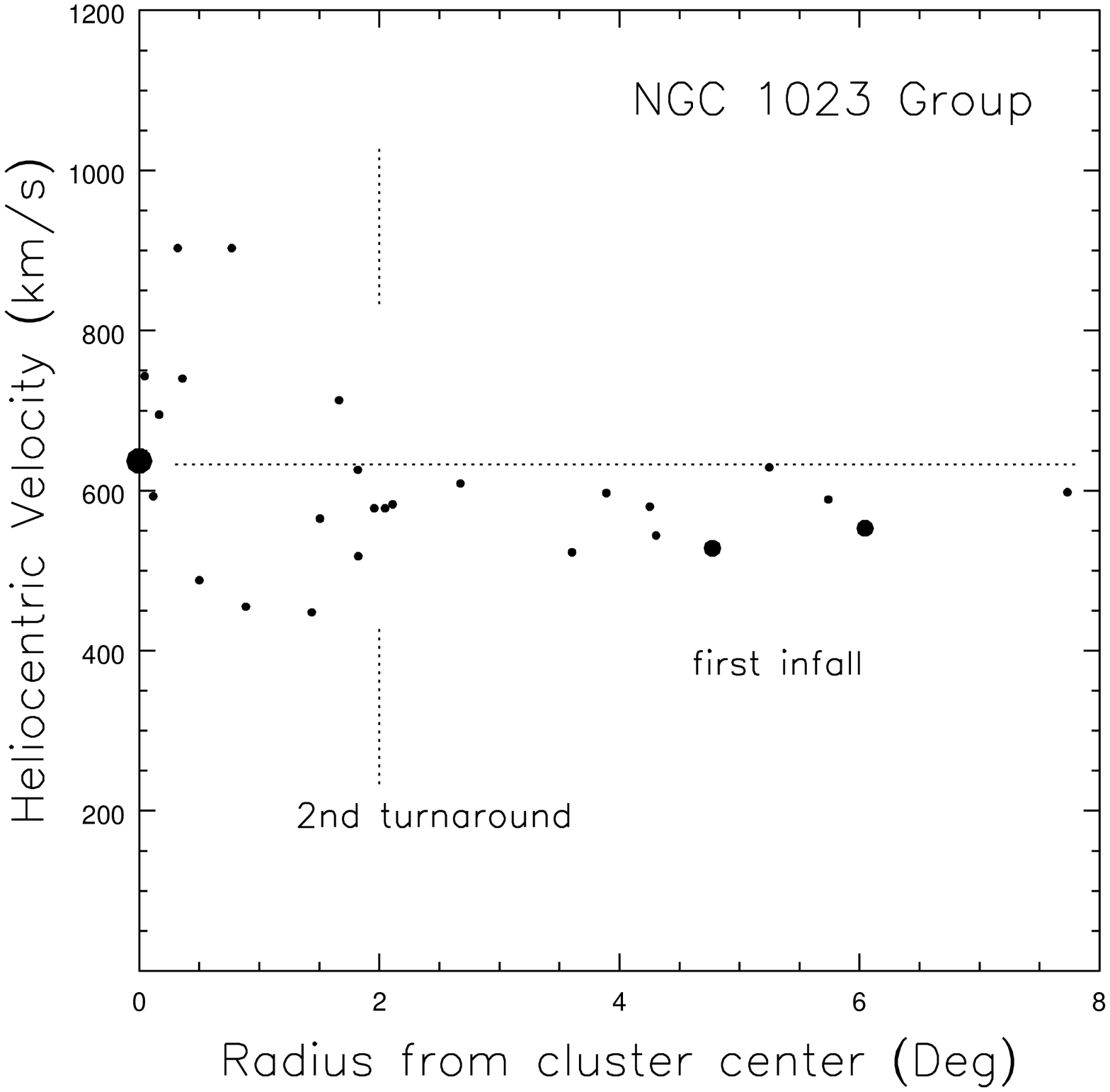, width=8.65cm}
\end{center}
\vskip-3mm
\caption{
Velocities of galaxies as a function of projected distance from NGC 1023, identified by the large symbol at radius zero.  The intermediate-sized symbols 
identify NGC 891 and NGC 925.  The radius of second turnaround centered on NGC 1023 lies at about $2^{\circ} = 350$~kpc.  Galaxies outside this radius are 
expected to be within the infall region around NGC~1023.
}
\label{rv}
\end{figure}

It is evident from Fig.~\ref{cand-all} and Fig.~\ref{cand-r2t} that candidate group members are strongly concentrated to the region around NGC~1023.  By analogy with what is known 
about other groups and 
clusters, it can be anticipated that galaxies in close proximity to NGC~1023 are in bound orbits within a massive halo while galaxies at
a large distance are infalling.  

This simple picture is supported by both velocity and density information.
The velocity information is presented in Figure~\ref{rv}.  Close to NGC~1023 velocity dispersions are 
high, but farther away they are moderate.  
A consistent picture emerges if $\sim 2^{\circ}$ is taken to be the radius about NGC~1023 of the surface of second 
turnaround of infalling galaxies
(see Mahdavi et al. 2005 for a definition and description of this parameter).  There are 17 galaxies with established velocities within $2.1^{\circ}=369$~kpc (with the distance 
to the group assumed to be
10 Mpc).  The mean heliocentric velocity of these 17 galaxies is $<V_h>=633$ km s$^{-1}$
with a dispersion of $\sigma_v=136$ km s$^{-1}$ and a standard deviation of 
33~km s$^{-1}$.  The harmonic 
mean (virial) radius is 315~kpc\ and the virial mass is $6.4 (\pm 3) \times 10^{12}$ M$_{\odot}$. The luminosity of the 41 galaxies within this
same radius around NGC~1023 is $L_R = 2.0 \times 10^{10}$ L$_{\odot}$ so 
the mass to light ratio is $M/L_R$ = 317~M$_{\odot}$/L$_{\odot}$.  These calculations follow the 
procedures of the earlier papers in this series.  As summarized in Tully \& Trentham (2008), the 
radius of second turnaround can be described by two alternative
relations:
\begin{equation}
r_{2t} = 0.193 (M_{12})^{1/3}~{\rm Mpc}
\label{r2t-mass}
\end{equation}
\begin{equation}
r_{2t} = \sigma_v /390~{\rm Mpc.}
\label{r2t-sig}
\end{equation}
From the derived values of $M_{12}$, 
the mass in units of $10^{12}$ M$_{\odot}$, and $\sigma_v$, the velocity dispersion, the NGC~1023 radius of second turnaround
from these two alternative relations is 358~kpc and 349~kpc, respectively.

\begin{figure}
\begin{center}
\vskip-4mm
\psfig{file=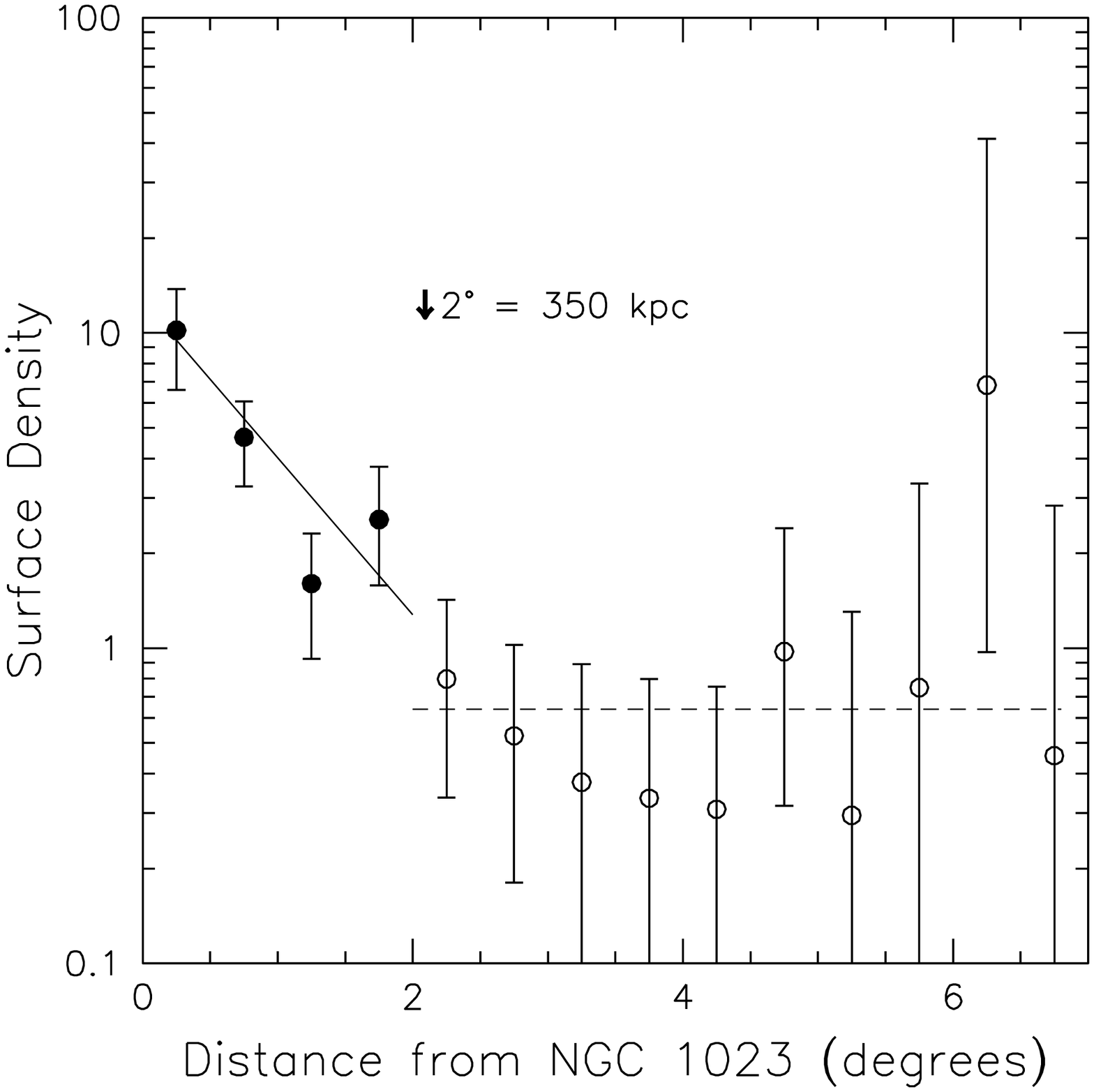, width=8.65cm}
\end{center}
\vskip-3mm
\caption{
Projected surface density of galaxies as a function of distance from NGC 1023.  The sample contains all galaxies in the survey region with known 
velocities less than 1000 km s$^{-1}$ and all of the 
candidates considered to be likely members.  Surface density falls off as $r^{-1/2}$ inside $2^{\circ}$ and is roughly constant outside this radius.
}
\label{rd}
\end{figure}

The fall-off in the surface density of galaxies with distance from NGC~1023 seen in Figure~\ref{rd} provides confirmation of a transition at roughly $2^{\circ}
$.  The discontinuities in velocity dispersion and density and the concurrence with expectations of Eqs. 4 and 5 give a consistent picture.  The second 
turnaround surface around NGC~1023 lies at about $2^{\circ} = 350$~kpc. Outside, the galaxies with known velocities in 
our sample have very close to the mean 
group velocity and must be bound, on 
first infall.  They have slightly lower velocities than the mean so it can be anticipated that these infalling galaxies are
slightly more distant than NGC~1023 and infalling with line-of-sight components toward us.

\begin{figure}
\begin{center}
\vskip-4mm
\psfig{file=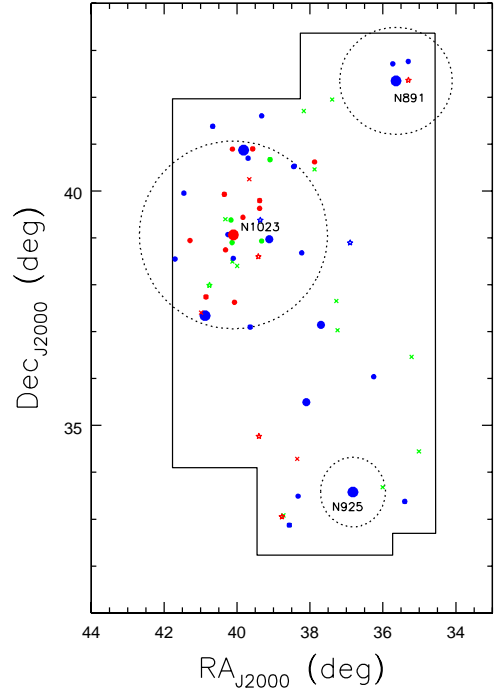, width=9.5cm}
\end{center}
\vskip-3mm
\caption{
Morphologies of the group candidates.  Types earlier than Sa are in red and later types are blue. Transition dE/dI and ambiguous types 
are green.  Filled circels: membership confirmed with a velocity or highly probable members.  Stars: suspected members brighter than 
$M_R = -11$.  Crosses: suspected members fainter than $M_R = -11$.  Dashed circles: projections of the surfaces of second turnaround for 
the three most luminous galaxies.
}
\label{early-late}
\end{figure}

\begin{figure}
\begin{center}
\vskip-4mm
\psfig{file=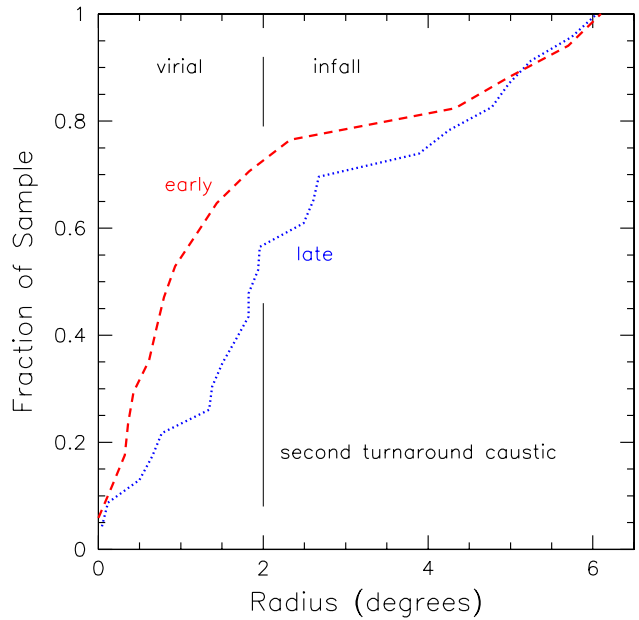, width=8.65cm}
\end{center}
\vskip-3mm
\caption{
Cumulative distributions as a function of distance from NGC 1023.  The sample of galaxies typed earlier than Sa has the behaviour 
described by the red dashed curve.  Galaxies later than Sa are characterized by the blue dotted curve.  Only galaxies with
$M_R < -11$ are considered. 
}
\label{cum_rad}
\end{figure}

Figure~\ref{early-late} adds further details.  All the confirmed and suspected candidates are represented.  Symbol shapes and colors indicate morphological 
types: early in red, late in blue, and transition or ambiguous in green.  There 
is an evident enhancement of the early types in close proximity to NGC~1023, 
within the $2^{\circ}$ circle.  The late types are more dispersed.  This visual impression is quantitatively confirmed with
Figure~\ref{cum_rad}.  Seventy--two percent of the early type galaxies within the MegaCam survey region lie within the second
turnaround radius and the median projected radius from NGC 1023 of these is $0.6^{\circ} = 105$ kpc.  By comparison, 55\% of late types lie within 
$r_{2t}$ and the median projected radius of these is $1.4^{\circ} = 245$ kpc.
The difference is marginal but there are enough galaxies that it is significant.

\begin{figure}
\begin{center}
\vskip-4mm
\psfig{file=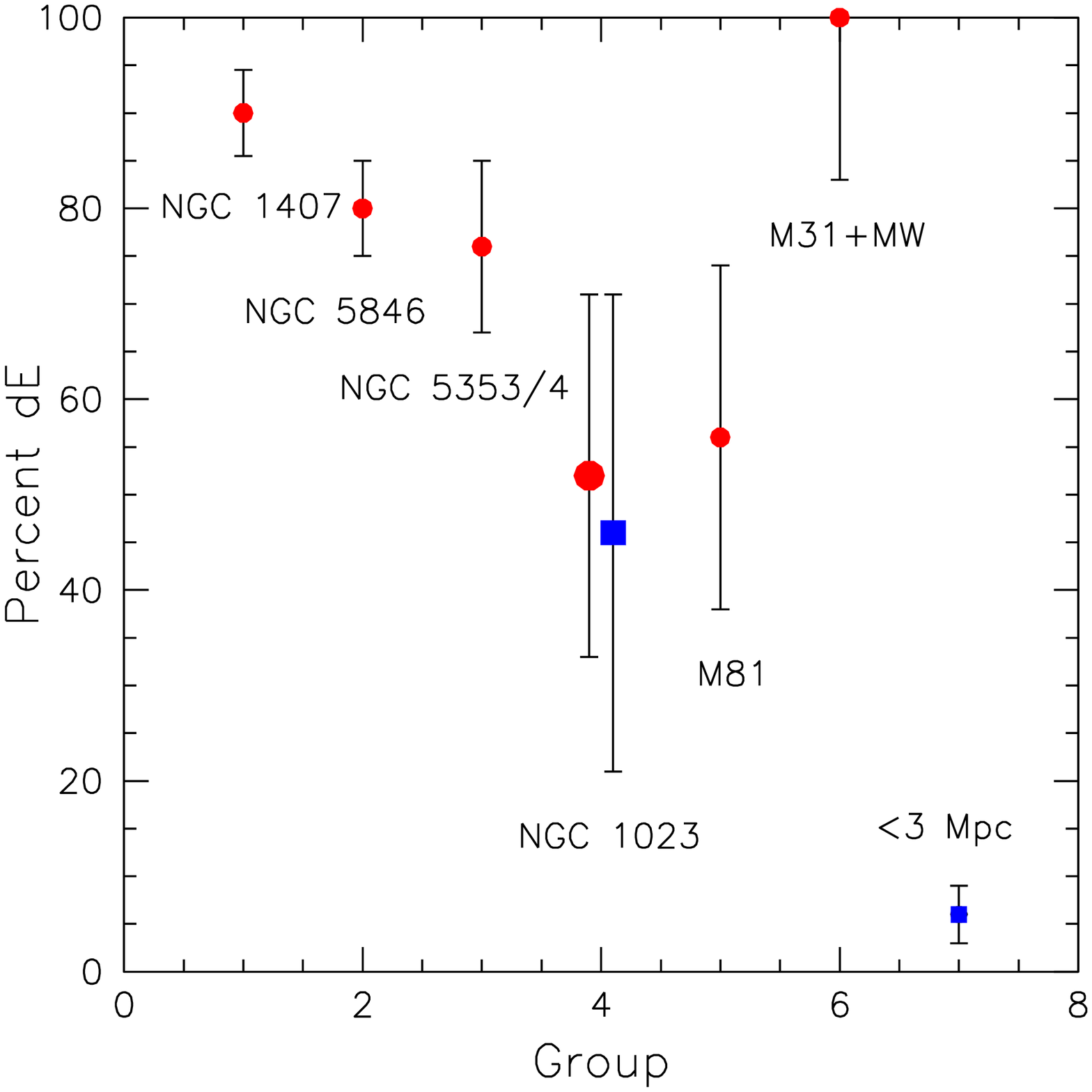, width=8.65cm}
\end{center}
\vskip-3mm
\caption{
Percent of dwarfs typed early in 6 groups and the field.  Galaxies are considered in the range $-17 < M_R < -11$.  Transition dE/dI and ambiguous cases 
are partitioned equally between early and late.  All the cases represented by red circles relate to regions within respective second turnaround 
radii.  The two cases represented by blue squares relate to regions outside of the infall caustic associated with a massive halo.
The x-axis is arbitrary and an indication of how unevolved each group is.}
\label{pcE}
\end{figure}

Circles are also plotted in Figure~\ref{early-late} around the second and third brightest galaxies, NGC~891 and NGC~925.  In the case of NGC~891, the HI linewidth implies a rotation velocity of 220 km s$^{-1}$ which 
suggests this system is similar in mass to the Milky Way, 
hence is expected to have a halo mass $\sim 1 \times 10^{12}$ and $r_{2t}$ dimension $\sim 200$~kpc = $
1.14^{\circ}$.  From its HI linewidth, NGC~925 rotates at 115~km s$^{-1}$ so would be a factor 3 less massive and have $r_{2t} \sim 130~{\rm kpc} = 0.74^{\circ}$.  It is seen that there 
are significantly fewer galaxies within either of these two domains than within the second turnaround radius of NGC 1023.

The tendency for dwarf galaxies to be dE rather than dI within the second turnaround radius is seen in Figure~\ref{pcE}.  Data from the earlier papers
in this series are brought forward.  In addition, there is information drawn from local neighborhood samples. Here, dwarfs are defined as galaxies 
with
$-17 < M_R < -11$.  The Local Group 
aside (where we have very good information but poor statistics), the lower mass, presumably less evolved places (NGC~1023 and
M81 groups) have lower fractions of early type dwarfs than the higher mass and more evolved environments.  Nonetheless, the early fractions are above half 
in all the second infall regions that have been observed in the course of this program.

\begin{figure}
\begin{center}
\vskip-4mm
\psfig{file=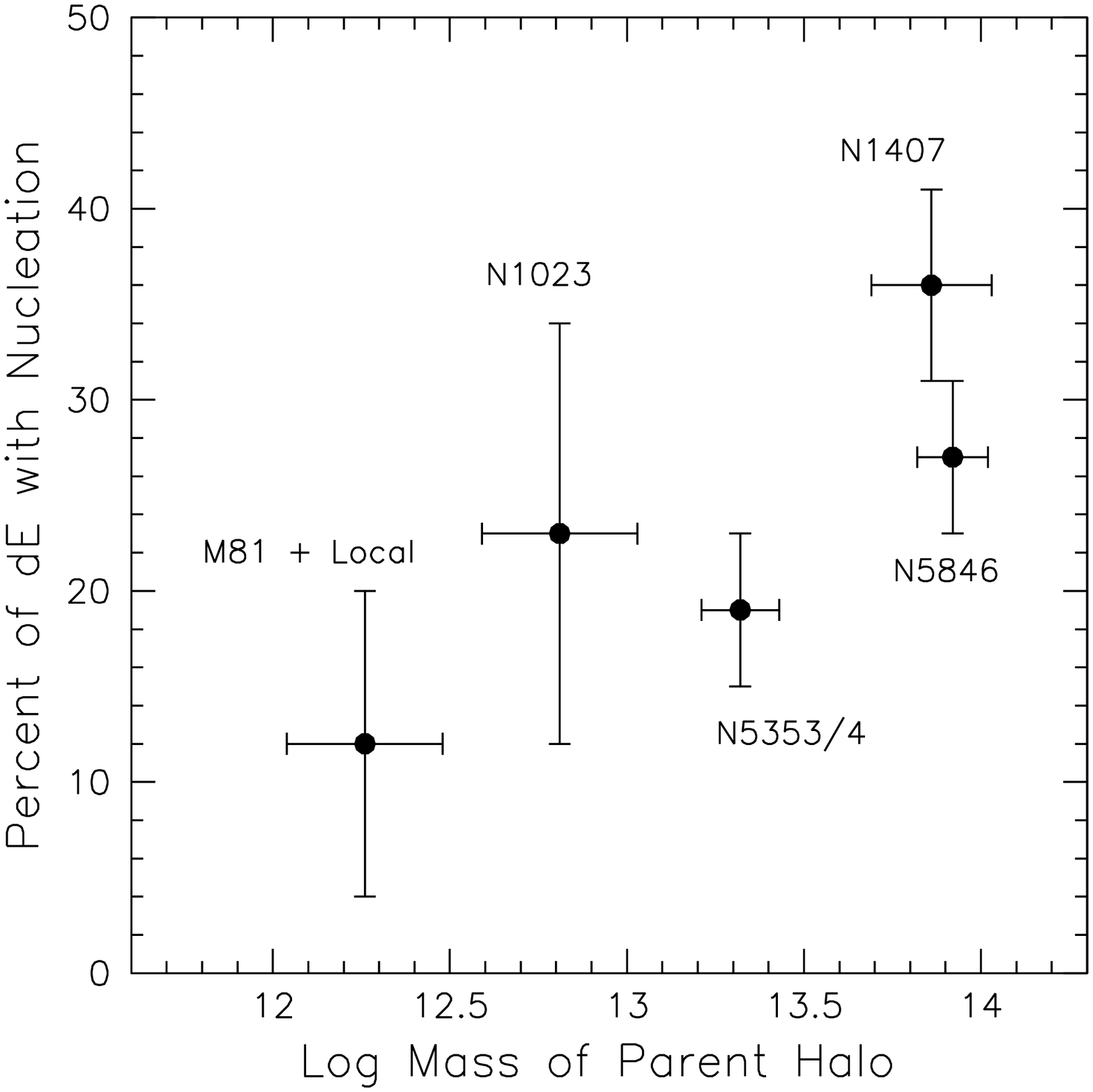, width=8.65cm}
\end{center}
\vskip-3mm
\caption{
The percent of dwarf ellipticals with nucleation (type dE,N) in the groups studied in this program.
Percentages are larger in the more massive and dynamically evolved halos.
}
\label{pcdEN}
\end{figure}

In some environments. many dwarf ellipticals have a central nucleation.  A substantial range of conditions have been probed over the series of papers
describing our CFHT MegaCam observations.  Morphological classifications dE,N (nucleated) and dE (non-nucleated) are given based on homogeneous image
material and consistent criteria.  Variations in the ratio of nucleated to non-nucleated dE are seen with Figure~\ref{pcdEN}.  The NGC 1407 and NGC 5846
groups are dense, dynamically evolved groups containing few late type systems.  As seen in Fig.~\ref{pcE}, most dwarfs in these groups are dE rather 
than dI.  Now as seen in Fig.~\ref{pcdEN}, roughly a third of these dE are nucleated.  The NGC 5353/4 Group is an intermediate case with a core of
early type galaxies, including a high fraction of dE rather than dI, but it is a group apparently receiving an influx of spiral and irregular systems.
The fraction of nucleated dE is lower.  The environments given attention here, around NGC 1023, M81, and the Local Group giants, are less evolved.
Though the statistics are poor, a trend is emerging of a lower fractional population of nucleated dE in less dense, less dynamically evolved regions.

\section{Luminosity function}

\begin{figure}
\begin{center}
\vskip-4mm
\psfig{file=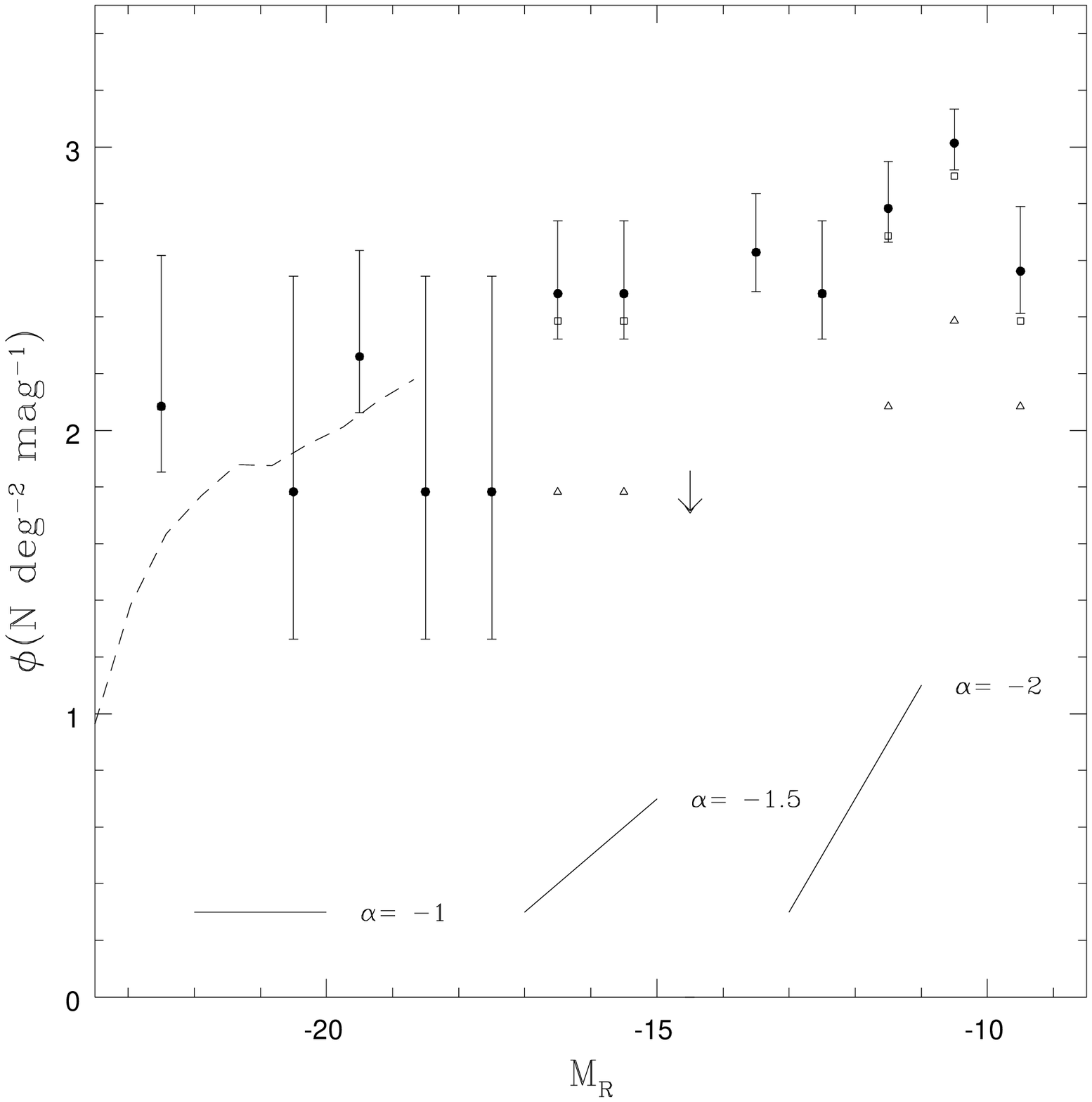, width=8.65cm}
\end{center}
\vskip-3mm
\caption{
The luminosity function of the NGC 1023 group. The contribution from galaxies within the second turnaround radius are shown as
open squares and the contribution from galaxies in the infall region are shown as open triangles.
Three values of $\alpha$, the logarithmic slope of the luminosity function, are shown.}
\label{lf_neil}
\end{figure}

\begin{figure}
\begin{center}
\vskip-4mm
\psfig{file=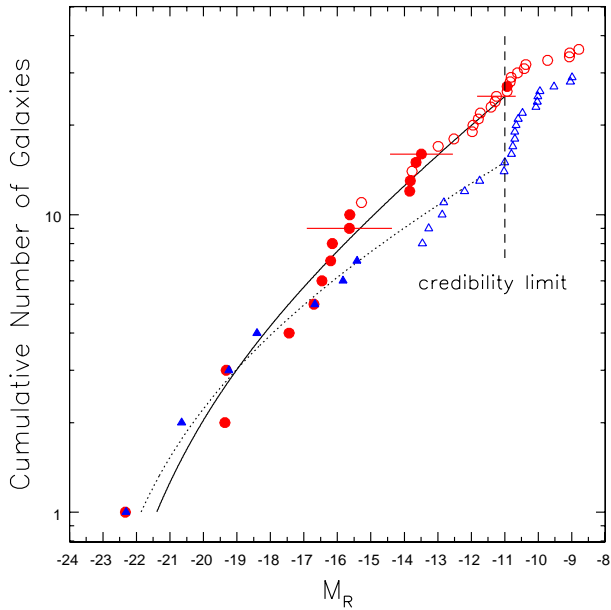, width=8.65cm}
\end{center}
\vskip-3mm
\caption{
Survey region cumulative luminosity functions.  Red circles: region within second turnaround radius about NGC 1023.  Blue 
triangles: radii beyond second turnaround; the first infall region.  Filled symbols: observed velocity confirms group membership.  
Open symbols: probable member.  Horizontal bars: typical uncertainties.  Solid curve: Schechter luminosity 
function fit to red circles, restricted to $M_R<-11$, with $\alpha=-1.22$ and $M^{\star}=-24.0$.  Dotted curve: fit to
blue triangles with $\alpha=-1.14$ and $M^{\star}=-25.5$.  Confusion/incompletion sets in fainter than $M_R=-11$.
}
\label{lf}
\end{figure}

The luminosity function for galaxies in the NGC 1023 survey region is presented in the standard differential form in Figure~\ref{lf_neil}.
It is seen that statistics even with 1~mag binning are desperately poor.  The cumulative luminosity function is more suited to a
situation with small numbers.  In Figure~\ref{lf}, one sees the separate cumulative luminosity functions for the regions within and
beyond the second turnaround caustic inferred to lie about NGC 1023.  

A `credibility limit' is indicated at $M_R = -11$.  It is seen
that the luminosity function flares upward at fainter magnitudes, especially in the sample beyond the second turnaround caustic. 
The surface density of this sample is sparse, creating an acute susceptibility to confusion from background contaminants.  The 
issue described earlier about contamination from dwarfs in the Perseus--Pisces background filament is evident.
Candidates at least two magnitudes fainter than $M_R = -11$ can be identified but they cannot reliably be distinguished from
dwarfs in the filament.

The situation brightward of $M_R = -11$ is reasonably secure.
Velocity measurements exist that confirm membership assignments for almost all candidates brighter than $M_R \sim -13.5$.
Candidates without known velocities but brighter than $M_R = -11$ are strongly clustered around NGC 1023 and a high fraction 
have morphological ratings 1 and 2, hence they are very probable members.  In summary, the 40 galaxies in the survey region with $M_R < -11$
are taken to be associated with the NGC 1023 Group, 25 of these within the second turnaround caustic and 15 within the
larger infall domain.  The least squares best fit Schechter (1976) luminosity functions for the two separate samples are 
illustrated by the solid and dotted curves in Fig.~\ref{lf}.

\begin{figure}
\begin{center}
\vskip-4mm
\psfig{file=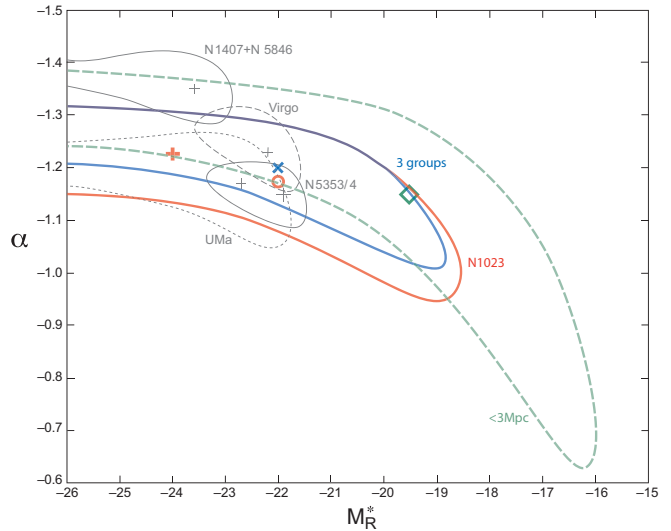, width=8.65cm}
\end{center}
\vskip-3mm
\caption{
Faint end slope parameter $\alpha$ vs. bright end cutoff parameter $M_R^{\star}$ of Schechter function fits to data for 8 groups and the field.  
The contour in red is the 95\% confidence limit for a fit to the galaxies in the collapsed region around NGC 1023.  The contours in blue
(labeled 3 groups) is the same limit for the sum of all galaxies in the collapsed regions around NGC 1023, M81, and in the Local Group.
The red cross and blue `X' indicate the locations of the $\chi^2$ minimizations for these two samples.  The open red circle indicates the location of best fit
with the NGC 1023 sample with the constraint $M_R^{\star} = -22$.   
The dashed green contour gives the 95\% probability limit for fits to a sample of 39 galaxies that lie within 3 Mpc but outside the M31 or Milky Way halos.
The bet fit for this field sample is indicated by the green diamond.
The fits and contours for the NGC 1407 + NGC 5846, 
Virgo, NGC 5353/4, and Ursa Major groups/clusters is extracted from an equivalent figure in Tully \& Trentham (2008).  If contours
extend beyond the left boundary of the plot, the data can be adequately fit with power laws without a high luminosity cutoff.
}
\label{mstar-alpha}
\end{figure}

There is a coupling between the Schechter parameters $\alpha$, describing the faint end power law, and $M_R^{\star}$, 
describing the bright end exponential cutoff.  The 95\% probability limits of the least squares fits are shown in
Figure~\ref{mstar-alpha}.  Give attention to the solid red contour labeled N1023.  The red cross locates the best fit
at $\alpha = -1.22$, $M_R^{\star} = -24$.  The contour is open at large negative $M_R^{\star}$.  A $\chi^2$ fit with $\alpha = -1.23$,
$M_R^{\star} = -\infty$ (a power law with no break) is negligibly worse than the optimal fit.  The counts are too small
to constrain a bright end cutoff.  From knowledge gleaned from better populated samples it can be anticipated that a cutoff is
appropriate.  If a cutoff with $M_R^{\star} = -22$ is specified then a fit only $0.2 \sigma$ worse than optimal is found 
with $\alpha = -1.17$.  The $1 \sigma$ uncertainty with the constrained fit is $\pm 0.05$.  The location of this 
preferred fit is shown by the red circle in Fig.~\ref{mstar-alpha}.

The M81 and Local groups provide samples that are useful for comparison.  Chiboucas et al. (2009) describe a related
program of observations with CFHT MegaCam of the second turnaround caustic region around M81.  Information pertinent
to the Local Group is gathered from the literature (Mateo 1998; van den Bergh 2006).  The Local Group is a dumbbell 
system with separate halos 
and caustics on scales of 200 kpc around M31 and the Milky Way and an infall region extending to a first turnaround
surface at $\sim 1$~Mpc (Karachentsev et al. 2009).

The blue contour in Fig.~\ref{mstar-alpha} labeled `3 groups' illustrates the 95\% probability limits of fits
to the sum of the NGC 1023, M81, and Local Group samples restricted in each case to the domains within the respective
second turnaround caustics.  The combined sample contains 63 galaxies with $M_R < -11$, 25 from the NGC 1023 Group,
23 from the M81 Group, and 15 from the Local Group.  A power law fit without a break is disfavored at the level of 
$1 \sigma$.  The value of $M_R^{\star}$ is poorly constrained but a best fit is found at $M_R^{\star} = -22$.
The slope $\alpha = -1.20 \pm 0.04$ is well constrained, though weakly coupled to the uncertain $M_R^{\star}$.

\begin{figure}
\begin{center}
\vskip-4mm
\psfig{file=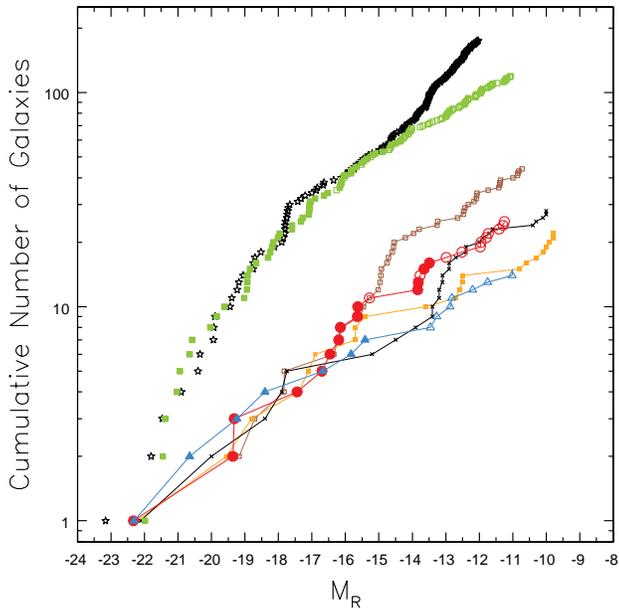, width=8.65cm}
\end{center}
\vskip-3mm
\caption{
Comparative cumulative luminosity functions.
Red circles and blue triangles: data transposed from Fig.~\ref{lf}.
Black stars: N1407 Group (Trentham et al. 2006).
Green squares: N5353/4 Group (Tully \& Trentham 2008).
Black crosses: M81 Group (Chiboucas et al. 2009).
Orange squares: M31 and Milky Way halos in Local Group.
Brown open squares: volume within 3 Mpc but excluding the halo regions around M31 and the Milky Way.
}
\label{lfplus}
\end{figure}

\begin{table*}
\caption{Properties of 6 Groups}
{\vskip 0.55mm} {$$\vbox{ \halign {\hfil #\hfil && \quad \hfil #\hfil \cr
\noalign{\hrule \medskip}
    & NGC 1407 &  NGC 5846 & NGC 5353/4 & NGC 1023 & M81 Group & Local Group &\cr
\noalign{\smallskip \hrule \smallskip}
\cr
Nearby Galaxies Catalog designation             &  51--8          & 41--1          & 42--1          & 17--1          & 14--10         & 14--12              &\cr
Distance (Mpc)                                  &   25            &   26           &   29           & 10.0           & 3.6            & $-$                 &\cr
No. ${\rm T} \leq 1$~$M_R<-19$ (early; bright)  &  13             &   11           &    5           & 1              & 0              & 0 $\dag$            &\cr
No. ${\rm T} > 1$~$M_R<-19$  (late; bright)     &   1             &     4          &    10          & 2              & 2              & 3 $\dag$            &\cr
$\sigma_V$ (km s$^{-1}$)                        & $387 \pm 65$    & $320 \pm 35$   & $205 \pm 28$   & $136 \pm 33$   & $110 \pm 25$   & $77 \pm 19~\ddag$   &\cr
$2^{nd}$ turnaround $r_{2t}$ (kpc)              & 900             & 840            & 530            & 350            & 270            & 220 $\ddag$         &\cr
$L_R$ ($10^{10}~L_{\odot}$)                     &   21            &  26            &  20            & 5.2            & 4.4            & 5.0 $\ddag$         &\cr
$M_T$ ($10^{12}~M_{\odot}$)                     & $73 \pm 27$     &  $84 \pm 20$   &  $21 \pm 5$    & $6.4 \pm 3$    & $2.2 \pm 1$    & $2.1 \pm 1~\ddag$   &\cr
$M/L_R$ ($M_{odot}/L_{odot}$)                   & $340 \pm 130$   & $320 \pm 80$   & $105 \pm 35$   & $124 \pm 60$   & $50 \pm 25$    & $38 \pm 20~\ddag$   &\cr
No. members ($M_R < -11$)                       & 240             & 250            &  126           & 25             & 23             & 15 $\dag$           &\cr
dwarf/giant ratio  ($>-17 / <-17$)              & $6.5 \pm 1.3$   & $7.3 \pm 0.7$  & $2.9 \pm 0.6$  & $5.2 \pm 2.8$  & $3.6 \pm 1.8$  & $1.5 \pm 0.8~\dag$  &\cr
Faint end slope $\alpha$                        & $-1.43 \pm 0.05$&$-1.34 \pm 0.08$&$-1.15 \pm 0.03$&$-1.22 \pm 0.05$&$-1.25 \pm 0.05$&$-1.13 \pm 0.04~\dag$&\cr
Cutoff $M_R^{\ast}$                             &   $-$           & $-24.0$        &  $-21.9$       & $-24.0$        & $-25.0$        & $-23.5~\dag$        &\cr
Brightest galaxy                                & $-23.16$        & $-22.49$       & $-22.30$       & $-22.33$       & $-22.11$       & $-22.48~\dag$       &\cr
Percent dE with Nucleation                      & $36 \pm 5\%$    & $27 \pm 4\%$   & $19 \pm 4\%$   & $23 \pm 11\%$  & $12 \pm 8\%$   & $\dag \dag$         &\cr 
\cr
\hline
\cr
$\dag$~~Includes M31 and MW virial regions & & & & & & &\cr
$\ddag$~~Only includes M31 virial region~~~~ & & & & & & &\cr
$\dag \dag$~~combined with M81 group  & & & & & & &\cr
\cr
  \hline
\noalign{\smallskip \hrule}
\noalign{\smallskip}\cr}}$$}
\end{table*}

The cumulative luminosity functions for these distinct samples are seen in Figure~\ref{lfplus}.  There are not
significant differences between the core regions around NGC 1023, M81, and the Local Group centers M31 and the Milky Way.
A curious deficiency of objects in the interval $-15<M_R<-14$ is seen in the ensemble sample.  Equivalent information 
from our earlier studies are brought across: for the NGC 5353/4 Group (Tully 
\& Trentham 2008) and for the NGC 1407 Group (Trentham et al. 2006).  Schechter function 95\% probability constraints for these samples 
are carried over to Fig.~\ref{mstar-alpha}.  See Table~3 for summary information on the various groups given attention 
to date with this program.

Almost exclusively, the environments that have been considered up to now in this discussion lie within massive halos limited by 
the radii of second turnaround for spherical collapse.  The exception is the more extended region around NGC 1023 that
we described as the infall domain.  That region alone is too limited to give useful statistics regarding the luminosity
function outside of massive collapsed halos.  In order to provide a better comparison, we turn to the immediate region 
around our Galaxy.  There is now reasonable completion brighter than $M_R = -11$ in a volume extending to 3 Mpc (Karachentsev
et al. 2004).  
This volume contains 39 known galaxies brighter than the magnitude limit once regions of 250 kpc radius around each of 
M31 and the Milky Way are excluded.
The 95\% probability limits of Schecter luminosity function fits for this sample are
included in Fig.~\ref{mstar-alpha}.  The optimal fit is with the parameters $\alpha = -1.15 \pm 0.07$ at $M_R^{\star} = -19.5$.
It is seen, though, that the constraints on $M_R^{\star}$ are weak.  $M_R^{\star} = -\infty$ is disfavored by only $1 \sigma$.
There is too little representation at bright magnitudes to constrain a cutoff.  Within the uncertainties, the luminosity function
for the volume outside collapsed cores resembles what is seen in the denser environments.  Certainly, though, as seen in 
Fig~\ref{pcE}, the galaxy morphologies are different.  Almost all dwarfs are gas--rich dI in the places that have not merged
with major ($>10^{12}~M_{\odot}$) halos.  

As a final point regarding luminosity functions, it can be seen from those shown in Fig.~\ref{lfplus} that there is no hint of 
a turndown within the limit $M_R = -10$.

\section{Discussion}

\begin{figure}
\begin{center}
\vskip-4mm
\psfig{file=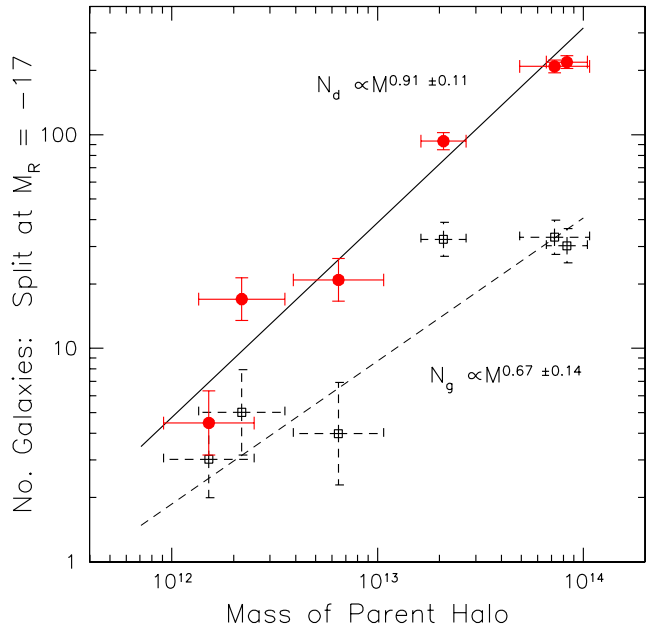, width=8.65cm}
\end{center}
\vskip-3mm
\caption{
Relationship between group halo mass and galaxy content.
The groups represented are, from right to left, NGC 5846, NGC 1407, NGC 5353/4, NGC 1023, M81, and Local Group.
The square black symbols identify the numbers 
of giant group members brighter than $M_R = -17$ while the round red symbols identify the numbers of dwarf group members in the interval $-17 < M_R < -11$.
In the case of the Local Group, an average is taken between the halos around M31 and around the Milky Way.
The solid straight line is a fit to the points representing the dwarf populations.  The dashed straight line is a fit to the
points representing the giant populations.
}
\label{massnum}
\end{figure}

A major goal when this series of papers began was to determine if there are significant variations in the 
luminosity function of galaxies with environment.  It was suspected that there were (Tully et al. 2002).
The admittedly poor evidence at the time suggested that denser, more dynamically evolved environments have
more dwarf galaxies and maybe steeper faint end slopes. 
The most evolved environments that we have studied, the NGC 1407 and NGC 5846 groups, have significantly steeper 
faint end slopes, with a joint best fit $\alpha = -1.35 \pm 0.03$.  It is seen in Fig.~\ref{mstar-alpha} that the 
95\% probability surfaces almost do not overlap with those of the other targets of this study.
The lowerer mass halo environments have an average value for the slope parameter of $\alpha = -1.23 \pm 0.04$
after normalizing to $M_R^{\star} = -23.5$ to give a fair comparison.
It can be seen in Table~3 that the two evolved groups have significantly the highest ratio of dwarfs per giant.  
However, there is Figure~\ref{massnum} to interpret.

Contributions to this plot come from the earlier studies in this series (of the NGC 5846, NGC 1407, NGC 5353/4, and M81 groups),
for the Local Group from the literature, and for the NGC 1023 Group with this work.  The parent halo masses
are derived from the velocities of group members through the virial theorem.  The ordinate records, alternatively, the number of
giant galaxies residing in the halo and the number of dwarfs.
It is seen that the number of giants ($M_R < -17$) is only loosly correlated with the halo mass while the number of dwarfs
($-17 < M_R < -11$) is quite tightly correlated.  The slope of a linear fit to the dwarf constituents is compatible with unity within the $1 \sigma$ uncertainty.
That is, the number of dwarf galaxies per unit halo mass is roughly constant.  Count the number of dwarfs and
one has a good estimate of the mass of the halo:
\begin{equation}
{\rm log} N_d = -10.2 (\pm 1.4) + 0.91 (\pm 0.11) {\rm log} M
\label{Nd-M}
\end{equation}
where $N_d$ is the number of dwarfs and $M$ is the mass of the halo in solar units.
We have established that there are statistically significant differences in luminosity functions in distinct environments
but the frequencies of dwarf galaxies normalized by the parent halo mass are statistically equal for these same environments.
The differences are in the giant galaxy populations.

Giant galaxy deficiencies are most convincingly seen in the NGC 1407 and NGC 5846 groups.  
These two groups are considerably more dynamicaly evolved than the other environments, to the degree in the case of
NGC 1407 that it almost qualifies as a `fossil group' (Ponman et al. 1994).
These groups contain one or two very bright members but a relative deficiency of intermediate luminosity giants.
It is expected that galaxy mergers have built up the dominant galaxies and depleted the reservoir of intermediate-sized systems
(Jones et al. 2003).

Although only weak differences have been found in the luminosity function faint end slopes in different environments 
there are strong differences in galaxy morphologies.
There is a dramatic difference in the ratio of dI to dE depending on whether one is considering a collapsed halo region 
(i.e. one in which the dark matter particles are bound on long timescales)
or a low density region outside of a large parent halo.
There is a more subtle dependence of the ratio of dE,N to dE, nucleation versus non-nucleation, on the degree of dynamic evolution of the host parent halo.

\section{Conclusions}

With the present study we have given consideration to a relatively low density environment, the sort of place where most gas-rich galaxies lie.
Over the course of a series of papers, we have discussed an interesting range of conditions in galaxy groups (UMa: Trentham et al. 2001;
NGC 5846: Mahdavi et al. 2005; NGC 1407: Trentham et al. 2006; NGC 5353/4: Tully \& Trentham 2008; M81: Chiboucas et al. 2009; NGC 1023: this paper).
The wide field and deep imaging with CFHT MegaCam has revealed the nature and extent of the regions of collapse in halos 
over the range $10^{12} - 10^{14}~M_{\odot}$.  
These regions can be characterized by the parameter $r_{2t}$, the radius of second turnaround with spherical collapse.

Almost everything that we have learned so far in this series pertains to the regions within the caustic delimited by $r_{2t}$.
Even in the present case of the NGC 1023 Group, which was chosen because all the known members save NGC 1023 itself are spirals or gas-rich irregulars,
a CCD mosaic survey has revealed that a swarm of dwarfs lies about the principal galaxy and a subsequent analysis of the pattern of velocities
tells us that there is a massive halo about that galaxy.

Interestingly, in this group there is a second galaxy, NGC 891, that is almost as luminous as NGC 1023 but this second most important galaxy 
does not have an appreciable entourage of dwarfs.
The early type S0 system NGC 1023, though barely more luminous than NGC 891, evidently lies within a much more massive halo.
A similar situation was found in the NGC 5353/4 Group studied earlier (Tully \& Trentham 2008).
The brightest galaxy in that group is NGC 5371 but this spiral has few dwarf companions.  
The large number of dwarfs congregate around the merging pair of S0 galaxies NGC 5353 and NGC 5354.

The most original discovery of the present paper is illustrated in Fig.~\ref{massnum}.
There is a tight correlation between halo mass and dwarf galaxy abundance.  
The slope of the correlation is consistent with a constant number of dwarfs per unit halo mass.
The implication is that the efficiency of dwarf production is {\it not} dependent on environment, at least across the spectrum of collapsed regimes we have explored.

Yet there are significant, although subtle, variations in luminosity functions.
The locations with few spirals, like the NGC 1407 and NGC 5846 groups, are relatively depleted in intermediate luminosity systems.
There is a suggestion of a dip at intermediate luminosities in the luminosity function of the NGC 5846 Group,
a feature seen in several rich clusters (Chiboucas \& Mateo 2009; McDonald et al. 2009).
An environment with spirals like the NGC 5353/4 Group retains (has not lost through merging?) its mid luminosity systems.  
The ratio of dwarfs to giants given in Table~3 provides a crude but robust characterization.

\begin{figure}
\begin{center}
\vskip-4mm
\psfig{file=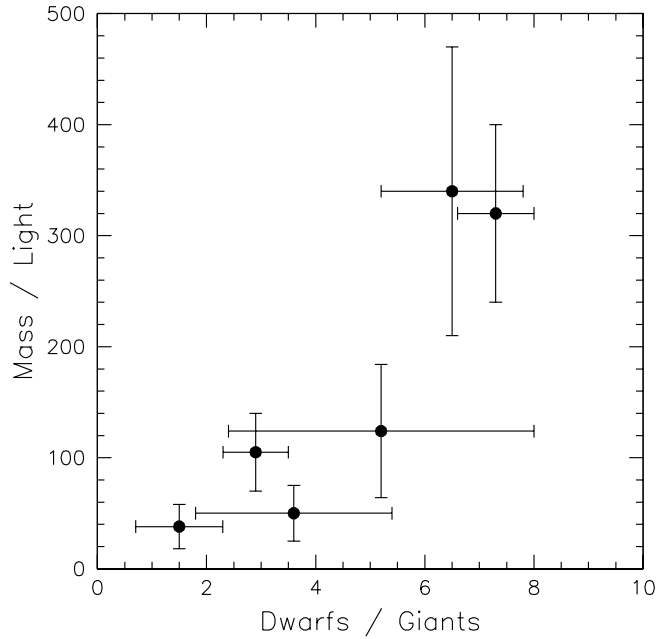, width=8.65cm}
\end{center}
\vskip-3mm
\caption{
The correlation between the ratio of dwarf ($-17 < M_R < -11$) to giant ($M_R < -17$) galaxies in groups with the virial mass to $R$ band light in those groups.
}
\label{d-g_m-l}
\end{figure}

The mass to light ratio is another characterization.
This ratio, also recorded in Table~3, is high in groups of dominantly early types and with high dwarf/giant ratios.
It is low in groups with substantial spiral fractions and low dwarf/giant ratios.  
The correlation between dwarf--giant ratio and mass--light ratio is shown in Figure~\ref{d-g_m-l}.
Those intermediate spirals are contributing both to a boost of the bright end of the luminosity function
and to a reduction of the mass to light ratio.
The correlation of mass to light ratio with parent halo mass was already demonstrated with a large sample by Tully (2005).

There is not yet sufficient information to quantify luminosity function differences between collapsed regions and the low density regions beyond second turnaround caustics.
We have given attention to the 3 Mpc region around the Local Group but it provides limited statistics.
Effectively by definition, a low density region outside a massive halo cannot contain a major galaxy, 
so a comparison of luminosity functions is meaningless at the bright end.
The coupling between the parameters $\alpha$ and $M^{\star}$ in the standard Schechter formulation creates an ambiguity in making comparisons.
The preliminary information shown in Fig.~\ref{mstar-alpha} suggests a similarity of faint end slopes between the field and groups.

Certainly, the morphologies of dwarfs are different outside of collapsed massive halos.  
Dwarfs ($-17 < M_R < -11$) are dE in the majority if found within the second turnaround caustic.  
The early types are inevitably more highly concentrated within the parent halo than late types.  
In the most dynamically evolved halos there is a greater propensity for dE to be nucleated.
By contrast, the dwarfs within 3 Mpc in the field are overwhelmingly late types.
Presumably gas-rich dwarfs that fall into a massive halo are transformed to dE.
There is evidence that such a transformation is happening today to dwarfs in the NGC 5353/4 Group.

It has become clear that the numbers of dwarfs are lower by a large factor from the numbers of 
low mass halos anticipated by Cold Dark Matter hierarchical clustering theory (Sheth \& Torman 1999).
As a point of comparison, if the faint end slope were $\alpha = -1.8$ with $M_R^{\star} = -22$ then
the ratio of dwarfs to giants would be 90.  

The observations of groups within the Local Supercluster reach limits of $M_R$ in the range --12 to --10.  
The inventories of group members may be incomplete for compact M32 type dwarfs in the more distant groups
but there is no indication that such objects make an important numeric contribution.
Dwarf galaxies exist but not in the huge numbers once anticipated by theory.
Dwarf galaxies will continue to be found with more sensitive surveys.  
There is no hint of a turndown in the luminosity function to a limit of $M_R = -10$.

\section*{Acknowledgements}
The major part of the observing program was performed at the Canada--France--Hawaii Telescope in queue mode with the wide field
imager MegaCam.  A substantial contribution to the initial analysis was made by Yannick Mellier and the Terapix team at the Institut
d'Astrophysique de Paris.  Spectroscopy was undertaken with Subaru Telescope with the collaboration of Andishai Mahdavi and Kristin
Chiboucas.  H\'el\`ene Courtois made an HI observation of a candidate with the Green Bank Telescope. 
RBT has been supported in his research with the US National Science Foundation grant AST 0307706.
This research has made use of the NASA/IPAC Extragalactic Database (NED)
which is operated by the Jet Propulsion Laboratory, Caltech, under agreement
with the National Aeronautics and Space Association.

\end{document}